\documentclass[10pt]{sig-alternate-10pt}
\usepackage{appendix}
\usepackage{caption}
\usepackage{color}
\usepackage{graphicx}
\usepackage{multicol}
\usepackage{multirow}
\usepackage{pseudocode}
\usepackage{verbatim}
\usepackage{subfigure}
\usepackage{url}
\usepackage{enumitem}

\usepackage{slashbox,pict2e}

\makeatletter
\def\sharedaffiliation{%
\end{tabular}
\begin{tabular}{c}}
\usepackage[final]{changes}
\definechangesauthor{F}{red} % Antonio
\definechangesauthor{A}{orange}
\definechangesauthor{J}{blue}
\definechangesauthor{V}{cyan}
\begin{document}

%\conferenceinfo{ACM e-Energy}{'13 Berkeley, CA, USA}

%COMMENTED BY ANGELOS ON 27/12/2013 
%\conferenceinfo{sigmetrics'14,} {June 16--20, 2014, Austin, Texas, USA.}
%\CopyrightYear{2014}
%\crdata{978-1-4503-2052-8/13/05}
%\clubpenalty=10000
%\widowpenalty = 10000
%COMMENTED BY ANGELOS ON 27/12/2013 

%\title{Measurements and Models of the Power Consumption of Data Center Servers}
%\title{Experimental analysis of the Power Consumption of Data Center Servers}
\title{A Measurement-based Analysis of the\\Energy Consumption of Data Center Servers\thanks{Partially
supported by Comunidad de Madrid grant S2009TIC-1692, MICINN grant TEC2011-29688-C02-01, 
and National Natural Science Foundation of China grant 61020106002.
The authors would like to thank Luis N\'{u}\~{n}ez Chiroque, Philippe Morere, and Miguel Pe\'on for their help with some experiments.}
}

%COMMENTED BY ANGELOS ON 27/12/2013 
\numberofauthors{4}
\author{
Jordi Arjona$^{*}$,  Angelos Chatzipapas$^{+,*}$, Antonio Fernandez Anta$^{+}$, and Vincenzo Mancuso$^{+,*}$\\
%	\alignauthor Jordi Arjona$^{*}$\\
%	\affaddr{jorge.arjona@imdea.org}\\
%	\affaddr{Institute IMDEA Networks, and University Carlos III of Madrid, Spain}\\
%	\alignauthor Angelos Chatzipapas$^{+,*}$\\
%	\affaddr{angelos.chatzipapas@imdea.org}\\
%	\affaddr{Institute IMDEA Networks, and University Carlos III of Madrid, Spain}\\
%\and
%	\alignauthor
%	\alignauthor Antonio Fernandez Anta$^{+}$\\
%	\affaddr{antonio.fernandez@imdea.org}\\
%	\affaddr{Institute IMDEA Networks, Madrid, Spain}\\
%	\alignauthor
%	\alignauthor Vincenzo Mancuso$^{+,*}$\\
%	\alignauthor
%	\affaddr{vincenzo.mancuso@imdea.org}\\
%	\affaddr{Institute IMDEA Networks, and University Carlos III of Madrid, Spain}\\
%	\alignauthor
%	\and
	\sharedaffiliation 
	\affaddr{$^*$University Carlos III of Madrid, Spain$\qquad\qquad\qquad\qquad^+$Institute IMDEA Networks, Madrid, Spain}\\
	\sharedaffiliation 
	\affaddr{\{jorge.arjona, angelos.chatzipapas, antonio.fernandez, vincenzo.mancuso\}@imdea.org}\\
%	\affaddr{Madrid, Spain}
	}
%COMMENTED BY ANGELOS ON 27/12/2013 

\maketitle
\begin{abstract}
\it
Energy consumption is a growing issue in data centers, impacting their economic viability and their public image. 
In this work we empirically characterize the power and energy consumed by 
different types of servers.
% Based on our measurements, we show how to predict and optimize 
%the energy consumed by an application. 
In particular, in order to understand the behavior of their energy and power consumption, we perform 
measurements in different servers.  In each of them, we exhaustively measure the power consumed by the CPU, 
the disk, and the network interface under different configurations, identifying the optimal operational levels. 
One interesting conclusion of our study is that the curve that defines the minimal CPU power as a function of the load 
%(measured in active cycles per second) 
is neither linear nor purely convex as has been previously assumed. Moreover, we find that 
the efficiency of the various server components can be maximized by tuning the CPU frequency and the number of active cores 
as a function of the system and network load, while the block size of I/O operations should be always maximized by applications.   
We also show how to estimate the energy consumed by an application 
as a function of some simple parameters, like the CPU load, and the disk and network activity. 
We validate the proposed approach by accurately 
estimating the energy of a map-reduce
%by means of the 
computation 
%of the {\it pagerank} metric of a graph 
in a {\it Hadoop} platform.

%New processors apply Dynamic Voltage and Frequency Scaling (DVFS) techniques. Since then, researchers have proposed various analytical models to estimate the power consumption of servers and thus to propose new methods for distributing the data. Moreover, there has been a lot of discussion whether new processors have linear or superlinear behavior in order to support their methods. In this paper we are the first to perform a complete power measurement characterization of the main components of a server, including CPU, memory, disk and network, using different frequencies under different load. Furthermore, we provide a simple analytical model to estimate the power consumption of a server taking into account frequency and load of the various components of the server. The results show that we can achieve significant power saving using the correct frequency for the CPU and that CPU power consumption has superlinear behavior.

\end{abstract}

\vspace{-2mm}
\category{\vspace{-1.5mm}B.8.2}{Performance and reliability}{Performance Analysis and Design Aids }[] 
.

\vspace{-3mm}
\terms{\vspace{-1.5mm}Measurement, Performance, Experimentation.}

%\keywords{\vspace{-1mm}DVFS; Network; Memory; Disk; Power Consumption}
\vspace{-3mm}
\keywords{\vspace{-2.5mm}Measurements; Power and Energy consumption; DVFS; Network; Disk I/O.}
\vspace{-1mm}

\section{Introduction}
\label{sec:introduction}

Massive data centers are becoming common nowadays. Large companies such as Google, Yahoo!, Amazon or Microsoft have deployed large data centers, housing tens of thousands of servers, and consuming a huge amount of energy every year. According to Van Heddeghem and Lambert~\cite{lambert},
%Koomey~\cite{koomey2011growth} 
%the estimation about 
data centers' total energy consumption in 
%2010 was about $240$~TWh, which corresponds to about $1.3\%$ of the global electricity consumption, and has an approximated annual growth rate of $9\%$ since 2005.
%in page 17 of the paper I take the mean value between upper and lower bound
$2012$ was about $270$ {\it TWh}, which corresponds to almost $2\%$ of the global electricity consumption, and has an approximated annual growth rate of $4.3\%$. 
This trend has driven researchers all over the world to focus on energy efficiency in data centers. 
Examples of energy saving techniques proposed during the recent years are virtualization plus consolidation and scheduling optimization \cite{kusic2009power, moore2005making}. 
%In a nutshell, virtualization consists of running tasks inside virtual machines, so that multiple tasks can be run independently in a server and migrated 
%between servers easily. Combined with virtualization, consolidation is the technique of aggregating virtual machines in servers to improve the amount 
%of resources used per server, and hence the efficiency of the whole data center. 
%Additionally, smart scheduling of tasks can also contribute to reduce energy consumption, by modulating the set of tasks that are run 
%simultaneously and deciding the order in which they have to be run \cite{moore2005making}. 
%Thanks to these techniques \cite{chase2001managing}, the energy efficiency of data centers has been dramatically improved in the last decade.
However, 
%although research has led to important advances towards energy efficiency, 
industry requirements keep increasing, and more research is necessary.

In this paper, out of all possible components of a data center, e.g., servers, routers, switches, etc., we concentrate on the characterization of servers and the
energy they consume. 
\deleted[V]{As mentioned, there has been already significant effort on devising energy saving techniques in servers \cite{??}. 
Examples of energy saving techniques proposed during the latest years are virtualization plus consolidation and scheduling optimization \cite{??}. 
In a nutshell, virtualization consists of running tasks inside virtual machines, so that multiple tasks can be run independently in a server and migrated 
between servers easily. Combined with virtualization, consolidation is the technique of aggregating virtual machines in servers to improve the amount 
of resources used per server, and hence the efficiency of the whole data center. 
Thanks to these techniques \cite{something}(cite showing benefits of virtualization/consolidation) the energy efficiency of data centers has been dramatically improved in the last decade. Additionally, a smart scheduling of tasks has also contributed to reduce energy consumption, by modulating the set of tasks that are run simultaneously and deciding the order in which they have to be run \cite{??}.}
Indeed, in order to obtain full benefit of the aforementioned energy-efficient techniques, it is crucial to have a good characterization 
of servers in the data center, as a function of the utilization of the server's components.
That is, it is necessary to know and understand the energy and power consumption of servers and how this changes under the different configurations. 
There is a large body of literature on characterizing servers' energy and power consumption. However, the existing literature does not jointly consider 
phenomena like the irruption of multicore servers and dynamic voltage and frequency scaling (DVFS)~\cite{weiser1996scheduling}, 
\added[V]{which are key to achieve scalability and flexibility in the architecture of a server}. 
With these new parameters, more variables come into play in a server configuration. 
Learning how to deal with these new parameters and how they interact with other variables is important since this may lead to larger savings.
% in the future. 
%as they do have an effect on technologies like the aforementioned virtualization, consolidation or scheduling.

It has been traditionally considered that the CPU is responsible for most of the power being consumed in a server, and that
this power increases linearly with the load. Although the power consumed by the CPU is significant, we believe that the power incurred by other elements of the server, like disks and NICs (Network Interface Cards) are not negligible, and have to be taken into account. Moreover, we believe that the assumption that CPU power consumption depends linearly from the load in a server may be too simplistic, especially when the server has multiple cores and may operate at multiple frequencies. In fact, even the way
load is expressed has to be carefully defined (e.g., it cannot be defined as a proportion of the maximal computational capacity of the CPU, since this value changes with the operational frequency).
Therefore, more complex/complete models for the power consumed by a server are necessary. 
In order to be consistent, these models have to be based on empirical values. 
However, we found that there is a lack of empirical work studying servers energy behavior.

%However, with the burst in of frequency scaling more variables have come into play and learning how to deal with them will lead us to larger savings in the future. One of the most important aspects in order to take full benefit of the aforementioned techniques is to know our machines and servers. In order to save we first have to know and understand. Nowadays, multicore computers are everywhere and frequency scaling does have an effect on how virtualization, consolidation and scheduling have to be performed in them.

%There is a large body of literature on energy efficiency and characterizing servers power consumption. In general, it is assumed that cpu is responsible for most of the power consumption in a server, however, more work in empirically characterizing server consumption is needed. Due to this lack of empirical works, research community is divided in $2$ major tendencies, depending on whether we assume that cpu power consumption has a linear or superlinear behavior with respect to the load in the server.

Our work tries to partially fill this void 
%of empirical analyses of server power consumption 
\added[V]{by proposing a measurement-based characterization---which is the first of its kind---of the energy consumption of a server with DVFS and multiple cores.} 
We evaluate here different server machines and evaluate what is the contribution to their power consumption of the CPU, hard drive disk, and network card (NIC). 
Our results support, for instance, our belief that more complex models than linear  are required for CPU power consumption. 
From the measurements obtained from the servers we evaluate, we propose a holistic energy consumption characterization, 
that accounts for the power consumed by CPU, disk, and NIC. Our approach captures the influence of the processing frequency and the multiple cores, not only to the 
CPU power consumption, but also to that of disk input/output (I/O) and NIC activity.

%
%Finally, we will see what are the consequences of using such a model when applying techniques such as consolidation or scheduling and how does it affect to some common assumptions related to linear models, as piling up as many virtual machines as possible in a server or keep servers running at their maximum speed.
%
\paragraph{Main results and contributions}
Our main contributions are of two kinds: $(i)$ we propose a methodology for empirically characterizing the energy consumption of a server, and $(ii)$ we provide novel insights on the power and energy consumption behavior of the most relevant server's components. 

As concerns the methodology, we observe that {\it active CPU cycles per second} (ACPS) is a convenient metric of CPU load in architectures using multiple frequencies and cores. We show how to isolate the contribution to energy/power consumption due to CPU, disk I/O operations, and network activity by just measuring the total server power consumption and a few activity indicators reported by the operating system. We also show that the {\it baseline} power consumption of a server---i.e., the power consumed just because the server is on---has a strong weight on the total server consumption. 
%This result enlightens the need of new efficient solutions for servers. 

As concerns the components' characterization, we show that, besides the {\it baseline component}, the CPU has the largest impact among all components, and its power consumption is not linear with the load. Disk I/O operations are the second highest cause of consumption, and their efficiency is strongly affected by the I/O block size used by the application. Eventually, network activity
plays a minor yet not negligible role in the energy/power consumption, and the network impact scales almost linearly with the network transmission rate. All other components can be accounted for in the {\it baseline} power consumption, which is subject to minor variations under different operational conditions.

\setlist[itemize]{leftmargin=*}

\newenvironment{packed_itemize}{
\vspace*{-0.5em}
\begin{itemize}[leftmargin=*]
  \setlength{\itemsep}{1pt}
  \setlength{\parskip}{0pt}
  \setlength{\parsep}{0pt}
}{\end{itemize}
\vspace*{-0.5em}
}

The main results of our campaign of measurements and analysis can be listed as follows:  
\begin{packed_itemize}
%	\item We empirically observe that active CPU cycles per second (ACPS) is a convenient metric of CPU load in architectures using multiple frequencies and cores; 
	\item The CPU consumption depends on the number of active cores, the CPU frequency, and the load (in ACPS units).   
		Our measurements confirm that the power consumption with a single active core at constant frequency can be closely approximated by a linear function of the load.
		However, given a CPU frequency, the power consumption 
		%cannot be described by a linear function of the load when more than one core is used; in fact, the power consumption 
		is a concave function of the load and can be approximated by a low-order polynomial.  
		The power consumption for a fixed load is, in general, minimized by using the highest number of cores and the lowest frequency at which the load can 
		be served. However, the minimum achievable power consumption is a piecewise concave function of the load. 
	\item The power consumed by hard disks for reading and writing depends on CPU frequency and I/O block sizes.
		Both reading and writing costs increase slightly with the CPU frequency. While the consumption due to reading is not affected by block size, 
		the power consumed when writing increases with the block size. The reading efficiency (expressed in MB/J) is barely affected by the CPU frequency, while 
		writing efficiency is a concave function of the block size since it boosts the throughput of writing until a saturation value is reached. 
		%Moreover, since increasing the I/O block size boosts the throughput of writing, writing efficiency significantly increases with the block size. 
		%Throughput and efficiency of reading operations do not change with the block size {\bf {CHECK!!}}.  	
	\item The power consumption and the efficiency of the NIC, both in transmission and reception, depends on the CPU frequency, the packet size, and the transmission rate. 
		%We observe that the transmission power is practically constant with the packet size, while it is affected by the CPU frequency.
		%While the packet size strongly impacts the achievable transmission rate, the efficiency of data transmission operations is practically constant with different packet sizes. 
		The efficiency of data transmission increases almost linearly with the transmission rate, with steeper slopes corresponding to lower CPU frequencies. 
		%For what concerns data reception, the power is affected by packet size and CPU frequency.
		Although a linear relation between transmission rate and efficiency holds for data reception as well, small packet sizes yield higher efficiency in reception. 	 
%	\item Characterize the power consumption of CPU, memory, HDD and network with respect to frequency \added[F]{and multiple cores}.
	\item Overall, we provide a holistic energy consumption model that only requires a few calibration parameters for every different server that we want to evaluate (a universal power model will be too simplistic and inaccurate). We validate our model by means of a server computing the 
{\it pagerank} metric of a graph in a {\it Hadoop} platform, with bulky network activity, and we found that the error due our energy estimates is below $7\%$. 

%	\item Change the traditional unit of measure of CPU load from \replaced[F]{percentage }{load $\%$} to cycles per second (CPS). {\color{blue} Propose a new metric to characterize the load of the CPU \-- active cycles per second (ACPS)}
%	\item Show that CPU consumption depends superlinearly on CPU load.
\end{packed_itemize}

%The rest of the paper is organized as follows. Section~\ref{sec:methodology} 
%describes and justifies the experiments we ran.
%Section~\ref{sec:measurements} presents servers' specifications and the 
%measurements performed for each of them for every single component that we
%evaluated. Sections~\ref{sec:model} and~\ref{sec:validation} are devoted to 
%model and validate the power consumption of the servers based on calibration
%parameters that we have found in earlier Sections. Implications are 
%discussed in Section~\ref{sec:discussion}. Section~\ref{sec:related} provides
%information about related works and finally Section~\ref{sec:conclusions}
%concludes the paper.

%!TEX root = ./eenergy2014.tex

\section{Methodology}
\label{sec:methodology}
%In this section we will, first, describe the different hardware used for our experiments. Afterwards we will enumerate the different benchmarks and tests ran for each one of the analyzed components.
%
%All the machines were under Linux, what allowed us to access the system registers in order to be able to collect important data, such as the active cycles of the system, and to be able to change some parameters, such as the operating frequency of the system. Similarly, most of our tests are based on Linux tools. 
%
%In general, we performed, at least, two different tests independently for each one of the evaluated components in order to verify that the obtained results were solid.

In this section we introduce the measurement techniques we used to characterize the power consumption of 
%the basic operations leading to non-constant power consumption in a server, i.e., 
CPU activity, disk access (read and write operations), and network activity.
\replaced[V]{Our measurements}{These experiments} start \replaced[J]{characterizing the CPU power consumption, from where we obtain information about the baseline power consumption of the system. After CPU and baseline characterization, we follow with experiments 
%we performed 
for the other two components,
% of a server that we analyzed, 
namely, disk and network.} 
{what we called the {\it baseline experiment} and follow with the different experiments we performed for each one of the analyzed components of a server, namely \replaced[V]{CPU, Disk and Network. }{CPU, HDD, network and memory.}}
Note that CPU and baseline measurements are of capital importance in order to evaluate the other components, because any operation run in a machine is like a puzzle with multiple pieces and we must know what is the contribution of each one of these pieces. \replaced[V]{Consider }{Think} that, we are paying a cost just for having a server switched on and the \replaced[V]{operating system }{OS } running on it. Similarly, every time we run a task in the system, some CPU cycles are needed in order to execute it as well as 
\replaced[V]{to use }{using } the component that has to perform \replaced[V]{the task}{it}. Hence, in order to understand the contribution of any component, we first need to \replaced[V]{identify }{detach} the contribution of the CPU and compute the difference \added[V]{with} respect to the aforementioned baseline.

%Although these experiments might seem to be independent one another they are intimately related. We first need to calculate the baseline of the system because it will tell us what is the cost of having an idle machine, so, when the experiments for CPU are run, the consumption due to the performed tasks will be the measured consumption minus the baseline. Similarly, the reason why we first evaluate CPU and not any other component is to be capable of characterizing the power consumed by the machine when running a certain amount of cycles per second. It is relevant because, when running a read(write) operation or when sending (receiving) data from the network, we are not only using the HDD or the NIC but also CPU in order to execute the commands that lead to these actions, then, to know what is the increase in the consumption due to the HDD or the NIC we first need to subtract the consumption due to the CPU. Only then we will know what is the contribution of each one of these elements separatedly.

\replaced[J]
{To explore the possible parameters determining the power consumption of a server and to gain statistic consistency we run our experiments multiple times. Similarly, we run these experiments in different servers and architectures in order to validate our results and give consistency to our conclusions.}{
To explore the possible parameters determining the power consumption of a server and to gain statistic consistency we run multiple different tests for each of the 
server components and run each experiment multiple times. Note that introducing multiple tests, e.g., by using different benchmarks, allows us to validate our results against the
measurement noise introduced by the adopted benchmark softwares.}

\subsection{Collecting system data and fixing frequency parameters}
\label{sub:meth:sdata}
%Something common to all our experiments was collecting the amount of active cycles the system used while performing the different tasks. All the experiments were performed under Linux OSs what allowed us to access to some system information in a very easy way.

One prerequisite for our experiments was having Linux machines due to the kind of commands and benchmarks we wanted to use and, mainly, because of the possibility of adding some kernel modules and utilities,\footnote{For instance cpufrequtils, acpi-cpufreq.} which \replaced[J]{allows us}{allow} to change CPU frequencies at will.
In a Linux system, CPU activity stats are constantly logged, so we can periodically read the core frequency and the number of {\it active} and {\it passive} CPU ticks at each core.\footnote{File \texttt{/proc/stat} reports the number of ticks since the computer started devoted to {\it user}, {\it niced} and {\it system} processes, waiting ({\it iowait}), processing interrupts (i.e., {\it irq} and {\it softirq}), and {\it idle}. In our experiments we count both waiting and idle ticks as {\it passive} ticks, while we denote the aggregated value of the rest of ticks as {\it active}.}
Once we have the number of ticks and the core frequency, since a tick represents a hundredth of second, cycles can be calculated as $100$ {\it ticks/frequency}. 
%In general, we will compute the amount of active and passive cycles during a certain time slot by computing the difference in ticks between its final and initial instant in time and obtaining the equivalent amount in cycles.

We use active cycles per second (ACPS) instead of CPU load percentage to characterize CPU load because 
the latter depends on the CPU frequency used, as the higher the frequency the more the work that can be processed. Hence, a percentage of load is not comparable when different frequencies are used, while the amount of ACPS that can be processed can be considered as an absolute magnitude.
In order to get and set information about the operative frequency of the system we used the 
\texttt{cpufrequtils} package.\footnote{\url{https://wiki.archlinux.org/index.php/CPU_Frequency_Scaling}}
%accessed the following registers:
%\texttt{/sys/devices/system/cpu/*/cpufreq/scaling\_cur\_freq} gives us the current operating frequency,
%\texttt{/sys/devices/system/cpu/*/cpufreq/scaling\_available\_frequencies} shows the valid system frequencies,
%\texttt{/sys/devices/system/cpu/cpuX/cpufreq/scaling\_min\_freq} fixes the minimum valid frequency and 
%\texttt{/sys/devices/system/cpu/cpuX/cpufreq/scaling\_max\_freq} fixes the maximum valid frequency\footnote{In registers 3 and 4 we wrote cpuX as the right command depends of which core frequency we want to modify, where X is the cpuID. If we had a 4 cores machine valid commands would use cpu0, cpu1, cpu2 or cpu3.}.
With those tools, we can monitor the CPU frequency at which the system works and assign different frequencies to the cores.
%lessen the power or increase the performance. 
However, to limit the number of possible combinations to characterize, we fix the frequency to be the same for all cores.

\begin{comment}
\subsection{Baseline}
\label{sub:meth:baseline}
The first experiment we ran in any machine was what we called the baseline experiment. The intention of this experiment is to measure the consumption of the system when it is running nothing but the OS and our own script. Similarly, we keep disks and memory slots connected but disconnect any network cable.

Our script registers the amount of active ticks consumed during time slots of $30$ seconds for each one of the different available frequencies in a machine. With this experiment we gain some insights in what is the average power consumed just by having a machine switched on in an almost idle state. Knowing this baseline will mainly help us to understand the range in which consumption varies when we run CPU experiments.
\end{comment}

\subsection{CPU}
\label{sub:meth:cpu}

In order to evaluate the CPU power consumption we prepared a script based on the benchmark application, 
namely \texttt{lookbusy}.\footnote{\url{http://www.devin.com/lookbusy}.}
Note that \texttt{lookbusy} allows us to load one or more CPU cores with the same load.%, from $1$ to $99\%$.
Our \texttt{lookbusy}-based experiment follows the next steps: we first fix the CPU frequency to the lowest possible frequency in the system; then we run 
\texttt{lookbusy} with fixed amount of load for one core during timeslots of $30$ seconds, starting with the maximum load and then decreasing the load gradually. 
%by $3\%$ steps until reaching a final $3\%$ load. 
After the last \texttt{lookbusy} run we measure the power consumed during an additional timeslot with \textit{no} \texttt{lookbusy} load offered. 
We register the active cycles and the power used during each timeslot.

\replaced[J]{After taking these different samples for one frequency we move to the immediately higher frequency (we can list and change frequencies thanks to \texttt{cpufrequtils}) and repeat the previous steps. After going through all the available frequencies, we restart the whole process but increasing by one the number of active cores. We repeat this whole process until all the cores of the server are active. Note that when we change the frequency of the cores we change it in all of them, active or not, for consistency. Similarly, when we have more than one active core, the load for all the active cores will be the same.} 
{When running the \texttt{lookbusy} based experiment we follow the next steps: we first fix the current frequency to a certain value; in second place, we vary the load in one core from $100\%$ to $0\%$ 
\replaced[J]{progressively}{by $3\%$ steps in time slots of $30$ seconds} while we do not impose any load on the rest of cores; then we change the current frequency to the next available one 
\replaced[J]{(we can list them thanks to \texttt{cpufrequtils})}{(we know the available ones by using the registers presented in subsection \ref{sub:meth:sdata})}
 and repeat the process for each one of the available frequencies. Finally,
 \replaced[J]{we repeat this experiment by sequentially increasing the active number of cores up to the maximum available in the server under study.}
 {we will sequentially increase the number of active cores from $1$ to the maximum number of available cores and repeat the whole process for each number of cores.}
 For instance, in a machine with $10$ available frequencies and $4$ cores \replaced[J]{and assuming $3\%$ steps}{} we would need $100/3\cdot 10 \cdot 4$ time slots of $30$ seconds.}

Once explained the scheme of our experiments, we must clarify the meaning of running a timeslot with \textit{no} load. Note that zero-load is clearly not possible 
as there is always going to be load in the system due to, e.g.,  the operating system. However, during the timeslot in which we do not run \texttt{lookbusy}, we
measure the power corresponding to the operational conditions which are as close as possible to the ones of an idle system. Moreover, the decision of using timeslots of $30$ seconds
is to guarantee enough, yet not excessive, time for the measurements. In fact, as we start and stop \texttt{lookbusy} at the beginning and end of the timeslots, we need to ignore the first and the last few seconds of measurements in each timeslot to avoid measurement noise due to power ramps and operational transitions. 

The measured values of load (in ACPS) and power in each timeslot
are used to obtain a least squares polynomial fittings curve. These fittings characterize the CPU power consumption for each combination of frequency and number of active cores. We will use as \textit{baseline power consumption} of each one of these configurations the zero-order coefficient of the polynomial of these fittings
curves.

\subsection{Disks}
\label{sub:meth:disks}

The power consumption of the hard drive was evaluated using $2$ different scripts (for reading and writing) based on the \texttt{dd} linux command.\footnote{\url{http://linux.die.net/man/1/dd}.}
We chose \texttt{dd} as it allows us to read files, write files from scratch, control the size of the blocks we write (read), control the amount of blocks written (read) and force the commit of writing operations after each block in order to reduce the effect of operating system caches and memory. We combine this tool with flushing the RAM and caches after each reading experiment.

\replaced[J]{In both our scripts we perform write (read) operations for a set of different I/O block sizes and for different data volumes to be written (read). In each case we 
record the CPU active cycles, the total power and time consumed in each one of these operations for each combination of block size and available frequency.}
{We devised $2$ different benchmark scripts, for reading and writing respectively.
We defined, in both scripts, different block sizes and the amount of blocks of each size to be written (read). In each one the scripts we will register the time and energy consumed when writing (reading) with each combination of block size and frequency, i.e., considering $5$ different block sizes and a machine with $10$ frequencies the execution of each one of the scripts would consist of $5\cdot 10$ runs. It is important to remark that, after each writing (reading) operation, we will flush the memory buffers to avoid that the next operation is speeded up by using RAM memory.}

%We devised $2$ different benchmark scripts with \textbf{dd}, for evaluating both the costs of reading and writing operations independently. In both cases we used different block sizes (100Mb, 10Mb, 1MB, 100Kb and 10Kb), i.e., the size of the read blocks when performing a reading operation or the size of the committed blocks when writing; and registered the required time, the cpu cycles and the average power consumption.
%
%The size of the read (written) files was of $3$GB (1GB for block sizes of $100$ and $10$ KB) when reading and $1.5$GB (50MB for block sizes of $100$ and $10$ KB) when writing. Each experiment was ran for each one of the available frequencies in the server.
%
%With \textbf{cp} we tested \ldots

Finally, 
%thanks to the count of consumed active cycles, 
we identify the contribution of the hard drive to the total power consumption by subtracting the contribution of both the 
baseline and the CPU consumption from the measured total power.

Disk I/O experiments shed light on the relevance of the block sizes when reading or writing as well as whether there is an influence of the frequency on these operations.

\subsection{Network}
\label{sub:meth:network}
\replaced[J]{In order to evaluate the contribution of the network to the power consumption of a server, we devised a set of experiments based 
on the \texttt{iperf}\footnote{\url{http://iperf.fr/}} tool as well as on our own UPD-client-server C script.}{, 
Network experiments were performed using four different testing tools, i.e. \texttt{iperf}
%~\footnote{\url{http://iperf.fr/}}, 
\texttt{mgen}
%~\footnote{\url{http://www.nrl.navy.mil/itd/ncs/products/mgen}}
and two UDP traffic generators that we created to evaluate the results andvvverify the results \-- one is using \textsl{clock\_nanosleep()} and the other is using a \textsl{for loop} to regulate the traffic intensity.}
%
%{\color{blue} do we need to analyze all of them? For the moment I report only the last one}
\begin{comment}
For the network loading we use a simple UDP client-server application written in C. The
network load is controlled using a for-loop in the client side that increases processing
and delays the packet transmission. Inside the for-loop
we include a modulo operation (because some compilers are clever enough to understand if there
is some work to be done inside the for-loop and if not they skip the for-loop) and every time
that the modulo operation results to zero the client sends a packet to the server. 
Algorithm~\ref{alg:loop} describes the traffic control that we described earlier.

\begin{pseudocode}[ruled]{LoadControl}{load, pktSize}
%  \caption{Load control loop.}
	\label{alg:loop}
%  \begin{algorithmic}[1]
    \PROCEDURE{LoadControl}{load, pktSize}%\Comment{Any comment}
    totalPkts \GETS load / (pktSize*8) \\
	\WHILE load \leq maxLoad \DO
	\BEGIN
	  loopController \GETS 1200 * \\ (1000/load) * (pktSize/1470); \\
      \FOR k \GETS 1 \TO totalPkts \DO
      \BEGIN
      	\FOR j \GETS 1 \TO loopController \DO
      	  \BEGIN
      	  \IF j \pmod{loopController} = 0 \THEN
      	    \BEGIN
      	    SendPacket();
      	    \END
      	  \END
      \END
    \END
    %\EndFor
    \ENDPROCEDURE
  %\end{algorithmic}
\end{pseudocode}

\end{comment}

\replaced[J]{There are several aspects that we consider relevant in order to characterize the impact of the NIC on the total power consumption of a server and that led us to choose these two tools. The first is the ability of performing tests where the computer under study acts as a server (sender) or as a client (receiver) of the communication, in order to observe its behavior when sending data or receiving it. For the sake of clarity, we will use, from now on, the terms \textit{sender}, for the server injecting
traffic to the network, and \textit{receiver} for the server accepting traffic from the network.}
{To evaluate the impact of NIC on the servers, we perform power and activity measurements for both 
\textbf{sender} and \textbf{receiver} side. \textbf{Sender} is defined as the 
device that injects
traffic to the network and \textbf{receiver} is defined as the device that 
accepts the traffic}
\replaced[J]{The second aspect consists in the ability to change several parameters that we consider relevant for this characterization, namely, the packet size and the offered load, jointly with the frequency of the system.}{For our network experiments we 
test the impact of three different parameters to the activity and power consumption, i.e. the 
frequency of the CPU, the network load and the packet size. Furthermore, since we are interested
on the impact of the load on the network we only test UDP traffic. For consistency, each
experiment is realized three times and we report the average of the measurements.}

Our experiments consist, then, on measuring the data rate achieved, the CPU active cycles and the total power consumption of the server 
under study acting as sender or receiver
when using different packet sizes and different rates. 
We run each experiment multiple times for statistical consistency.

\replaced[J]{Finally, in order to isolate the consumption from the network, we characterize with the CPU active cycles measured in the experiment the consumption due to the CPU and the baseline and subtract them from our measurements.}
{To evaluate the measurements we get from the power meter, we substract the baseline power
consumption of the server (Sec.~\ref{sub:meth:baseline}), and afterwards using the activity measurements we substract the
power consumption that corresponds to the frequency used and the amount of active cycles 
(Sec.~\ref{sub:meth:cpu}), resulting the network power consumption.}

\section{Measurements}
\label{sec:measurements}

\subsection{Devices and Setup}

\replaced[J]{In order to monitor and store the instantaneous power consumed by a server during the different experiments we used a Voltech PM1000+ power 
analyzer,\footnote{\url{http://www.farnell.com/datasheets/320316.pdf}} which is able to measure the total instantaneous power consumed by the server under test on 
a per-second basis. In order to take our measurements we connected the server being measured to the power analyzer and the latter to the power supply.}
{In order to monitor and store the instantaneous power consumed by a server during the different experiments, we have used a power meter. The measuring device used has been a Voltech PM1000+ power analyzer inserted between the 
electricity plug of the wall and the server. This device is able to measure and store the instantaneous power consumed by the server every second. Note that the obtained measurements are always of the total power consumed by the server. 
}
\replaced[J]{In the experiments where the network was not involved (CPU and disk), we unplugged the network cable from the server, which has an impact on the power consumption as the port goes idle. In the network based experiments we established an Ethernet connection between the server under study and a second machine in order to study the server behavior, both as a receiver as well as as a sender.}
{In the network experiments, we connect via Ethernet
the server under evaluation with another computer, which depending on the experiment will act as sender or receiver of network traffic. 
For the experiments in which the
network is not involved, the network cable is physically disconnected from the server under evaluation (which in fact has a impact in the power consumed).} 

\begin{table}[tb]
\caption{Characteristics of the servers under study}
\vspace{-3mm}
    \label{tab:servers}
    \small
    \centering
        \begin{tabular}{|c|p{1.5cm}|p{1.5cm}|p{1.7cm}|}
                \hline
                     \multirow{2}{*}{Component}& \multicolumn{3}{|c|}{\textbf{Servers}}\\
                \cline{2-4}
                    & \texttt{Survivor} & \texttt{Nemesis} & \texttt{Erdos}\\
                \hline
                    CPU (\# cores) & 4 & $4$ & $64$ \\
                \hline
                    \# freqs &  8 & 11 & $5$ \\
                \hline
                    \multirow{11}{*}{Freqs List} & $1.2$~\textit{GHz}, 1.333~\textit{GHz}, 1.467~\textit{GHz}, 1.6~\textit{GHz}, 1.733~\textit{GHz}, 1.867~\textit{GHz}, 2~\textit{GHz}, $2.133\;$ \textit{GHz}
                     & 1.596~\textit{GHz}, 1.729~\textit{GHz}, 1.862~\textit{GHz}, 1.995~\textit{GHz}, 2.128~\textit{GHz}, 2.261~\textit{GHz}, 2.394~\textit{GHz}, 2.527~\textit{GHz}, 2.666~\textit{GHz}, 2.793~\textit{GHz}, $2.794\;$ \textit{GHz}
                      & 1.4~\textit{GHz}, 1.6~\textit{GHz}, 1.8~\textit{GHz}, 2.1~\textit{GHz}, $2.3\qquad$~\textit{GHz}\\
                \hline
                    RAM & $4$~\textit{GB} & $4$~\textit{GB} & $512$~\textit{GB} \\
                \hline
                    Disk & $2$~\textit{TB} & $2+3$~\textit{TB} & $2\times 146$\textit{GB} \\
                            &                         &                              & $4\times 1$~\textit{TB} \\
                \hline
                    Network & $1$~\textit{Gbps} & $3 \times 1$~\textit{Gbps} & $4\times 1$~\textit{Gbps}, $2\times 10$~\textit{Gbps} \\
                \hline
        \end{tabular}
        \vspace{-5mm}
\end{table}

\replaced[J]{We evaluated three different servers: \texttt{Survivor}, \texttt{Nemesis}, and \texttt{Erdos}. We will now present these servers although their main characteristics, including their sets of available CPU frequencies, can be also found in Table~\ref{tab:servers}.}
{We evaluate three different servers, which we will refer as \texttt{Survivor}, \texttt{Nemesis}, and \texttt{Erdos} in the rest of the document.}
\texttt{Survivor} has \replaced[J]{an Intel Xeon E5606 $4$-core processor, with}
{a Dell {\color{red} Optiplex 780 (CORRECT MODEL?)} with an Intel Xeon $2.133$~\textit{GHz} $4$-core processor,}
%with the following available frequencies: 1.2GHz, 1.333Ghz, 1.467GHz, 1.6GHz, 1.733GHz, 1.867GHz, 2GHz and 2.133GHz. It has
$4$~\textit{GB} of RAM, a $2$~\textit{TB} Seagate Barracuda XT hard drive and 
a $1$~\textit{Gigabit} Ethernet card integrated in the motherboard. 
\texttt{Nemesis} is a Dell Precision T3500 with an Intel Xeon W3530 $4$-core processor,
% the following available frequencies: 1.596GHz, 1.729Ghz, 1.862GHz, 1.995GHz, 2.128GHz, 2.261GHz, 2.394GHz, 2.527GHz, 2.666GHz, 2.793GHz and 2.794GHz. It has }
$4$~\textit{GB} of RAM, 2 hard drives (a $2$~\textit{TB} Seagate Barracuda XT and a
$3$~\textit{TB} Seagate Barracuda), a $1$~\textit{Gigabit} Ethernet card integrated in the
motherboard, and a separate Ethernet card with two $1$~\textit{Gigabit} ports. In this study we only evaluate the
Seagate Barracuda XT disk and the integrated Ethernet card.
Both \texttt{Survivor} and \texttt{Nemesis} 
%are regular desktop servers and 
use the Ubuntu Server edition 10.4 LTS Linux distribution.
Finally, \texttt{Erdos} is a Dell PowerEdge R815 with 4 AMD Opteron 6276 16-core processors (i.e., 64 cores in total),
% the following available frequencies: 1.4GHz, 1.6Ghz, 1.8GHz, 2.1GHz and 2.3GHz. It has
$512$~\textit{GB} of RAM, two $146$~\textit{GB} SAS hard drives configured as a single RAID1 system (which is the ``disk" analyzed here) and four $1$~\textit{TB} Near-line SAS hard drives.  \replaced[J]{It also includes four $1$~\textit{Gigabit} and two $10$~\textit{Gigabit} ports}{}. \texttt{Erdos} is a high-end server and uses Linux Debian 7 Wheezy.

\subsection{Baseline and CPU}
\label{sub:meas:cpu}

  \begin{figure}[!tb]
      \centering
      \subfigure[\texttt{Survivor}]{\label{fig:allCoresSurvivor}	\includegraphics[width=0.95\columnwidth, trim= 15mm 6mm 21mm 8mm, clip=true]{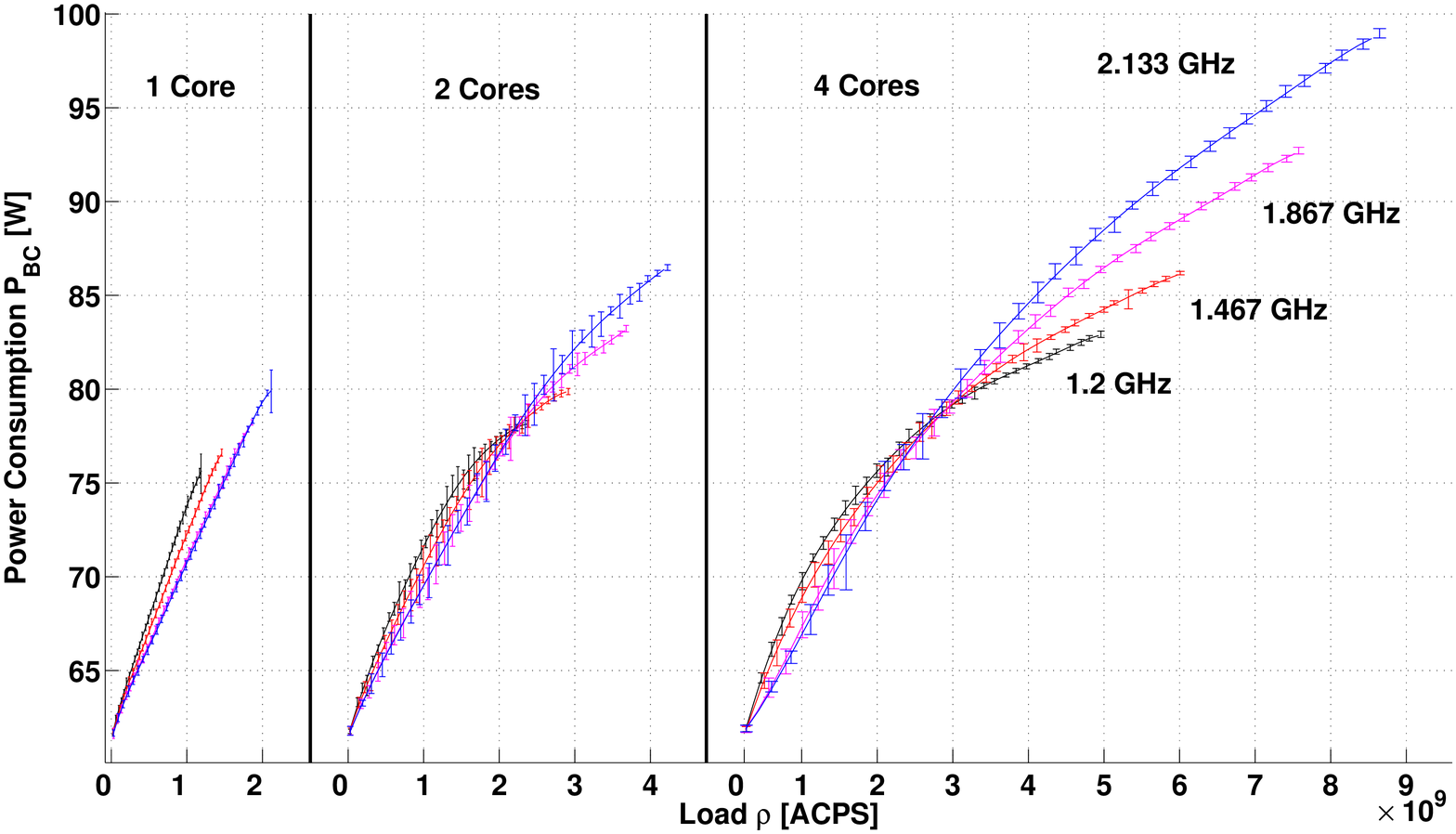}}
      \subfigure[\texttt{Nemesis}]{\label{fig:allCoresNemesis}	\includegraphics[width=0.95\columnwidth, trim= 15mm 6mm 21mm 0mm, clip=true]{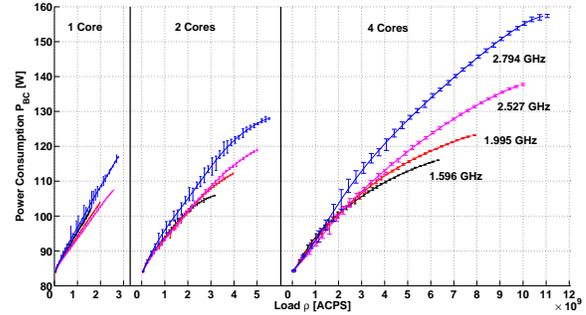}}
      \subfigure[\texttt{Erdos}]{\label{fig:allCoresErdos}	\includegraphics[width=0.95\columnwidth, trim= 24mm 6mm 21mm 0mm, clip=true]{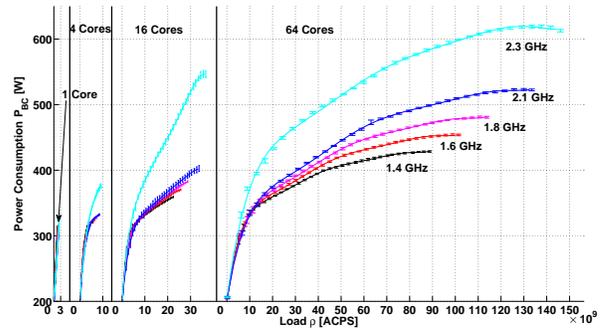}}
   	\vspace{-3mm}
   \caption{Power consumption of $3$ servers (\texttt{Survivor}, \texttt{Nemesis}, and \texttt{Erdos}) for baseline and CPU characterization experiments.}
     	\vspace{-3mm}
     \label{fig:allCores}
       \end{figure}

As mentioned in the previous section, for each server we have measured the power it consumes without disk accesses nor network traffic. We assume that the power consumption observed is the sum of the baseline consumption plus the power consumed by the CPU. We have obtained samples of the power consumed under different configurations that vary in the number of active cores used, the frequency at which the CPU operates (all cores operate at the same frequency), and the load of the active cores (all active cores are equally loaded).
The list of available and tested CPU frequencies and cores can be found in~Table \ref{tab:servers}.
We tune the total load $\rho$ by using \texttt{lookbusy}, as described in the previous section. Each experiment lasts 
$30$~{\it s} and it is repeated $10$ times. Results are summarized in terms of average and standard deviation. 
Specifically, in the figures reported in this section, the power consumption for each tested configuration is depicted by means of a vertical segment centered on the average power consumption measured, and with segment size equal to two times the standard deviation of the samples.  

   \begin{figure}[!tb]
      \centering
      \vspace{-2mm}
      \subfigure[Minimal power.]{\label{fig:nemesisEnv}\includegraphics[width=0.46\textwidth]{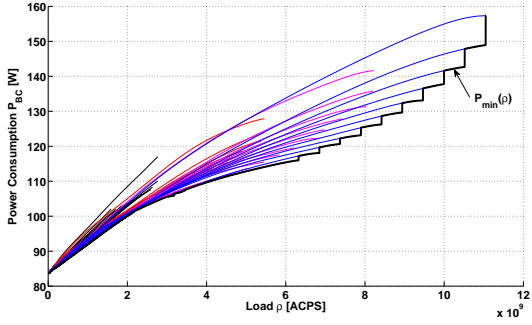}}\\
             \vspace{-4mm}
      \subfigure[Maximal efficiency.]{\label{fig:nemesisEff}\includegraphics[width=0.46\textwidth]{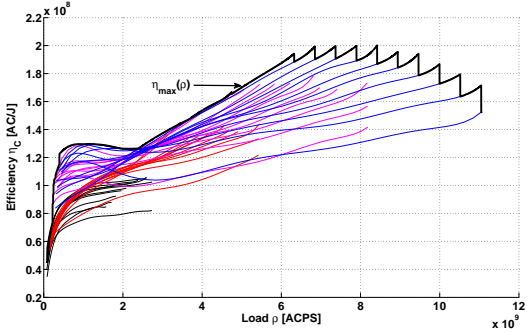}}
      \vspace{-4mm}
      \caption{CPU performance bounds of \texttt{Nemesis}.}
      \vspace{-6mm}
      \label{fig:nemesisenvelops}
    \end{figure}

  \begin{figure}[!t]
      \centering
      \subfigure[Minimal power.]{\label{fig:erdosEnv}\includegraphics[width=0.46\textwidth]{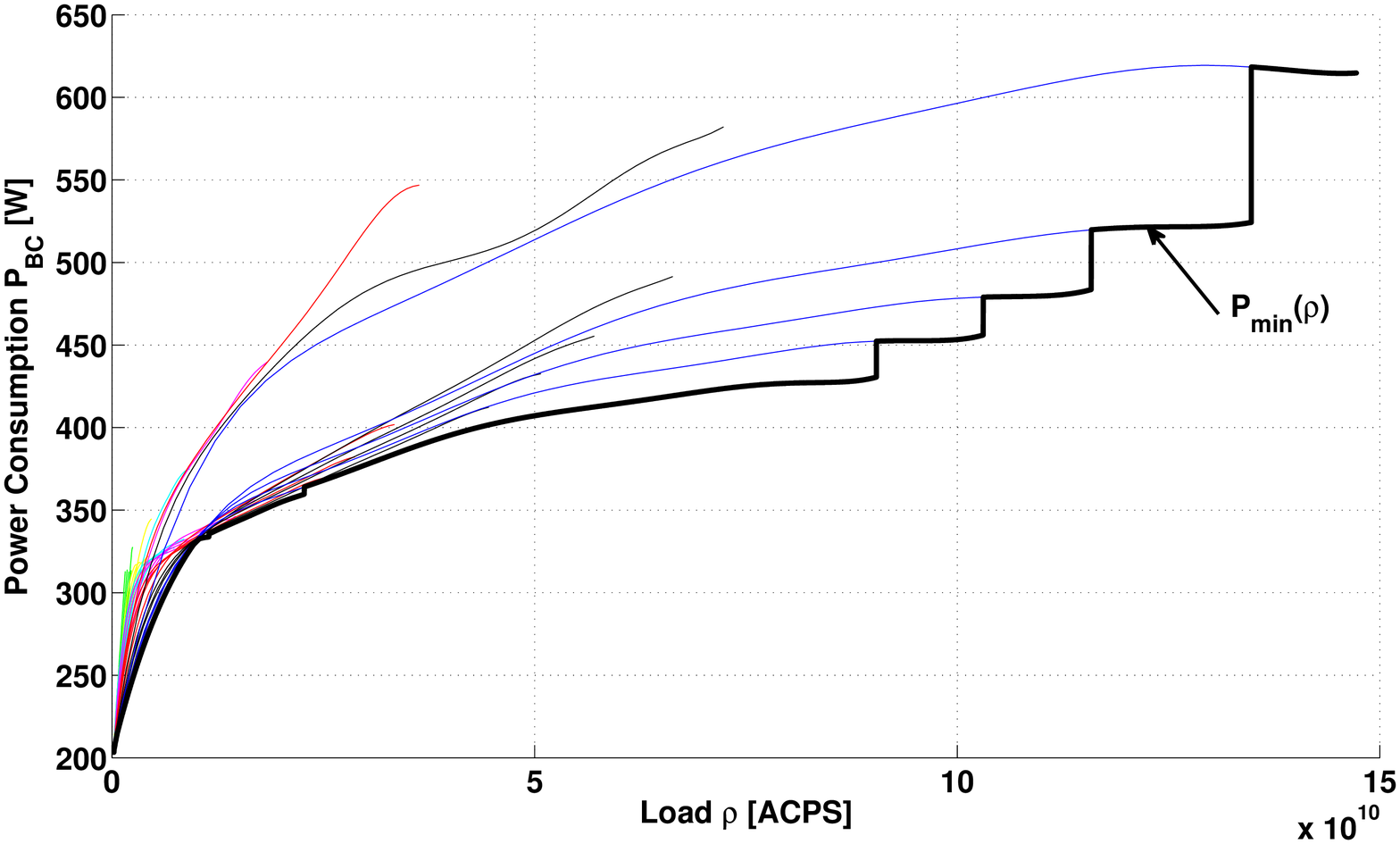}}\\
             \vspace{-4mm}
      \subfigure[Maximal efficiency.]{\label{fig:erdosEff}\includegraphics[width=0.46\textwidth]{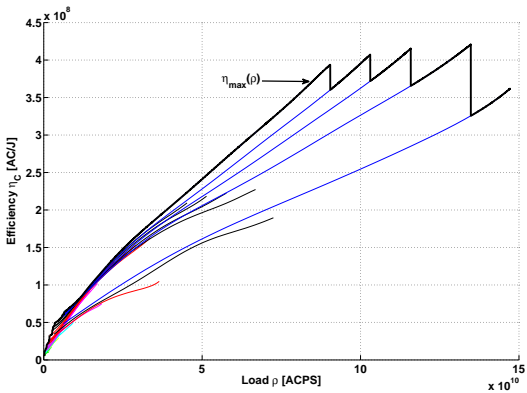}}
             \vspace{-2mm}
      \caption{CPU performance bounds of \texttt{Erdos}.}
            \vspace{-4mm}
      \label{fig:erdosenvelops}
    \end{figure}

The results of these experiments for each of the $3$ servers are presented in Figure~\ref{fig:allCores} (the measurements for some frequencies and some number of cores are omitted for clarity). Here, for each configuration of number of active cores, frequency, and load in ACPS, the mean and standard deviation of all the experiments with that configuration are presented. Also the least squares polynomial fitting curve for the samples is shown for each number of cores and frequency. The curves shown are for polynomials of degree $7$, but we observed that using a degree $3$ polynomial instead does not reduce drastically the quality of the fit (e.g., the relative average error of the fitting increases from $0.7\%$ with $7$-th degree polynomials to $1.5\%$ with degree equal to $3$ for \texttt{Erdos}, while it remains practically stable and below $0.7\%$ for \texttt{Nemesis}).
In general, we can use an expression like the following to characterize the CPU power consumption: 
\begin{equation}
\label{eq:PBC}
\centering
P_{BC}(\rho) = \sum_{k=0}^n \alpha_k \rho^k, \quad n \le 7,  
\end{equation}  
where $P_{BC}$ includes both the baseline power consumption of the servers and the power consumed by the CPU, and $\rho$ is the load expressed in active cycles per second. 
Therefore, coefficient $\alpha_0$ in Eq.~\ref{eq:PBC} represents the consumption of the system when the CPU activity tends to $0$, and we can thereby interpret $\alpha_0$ as the 
baseline power consumption of the system. Note that the polynomial fitting, and hence the baseline power consumption $\alpha_0$, depends on the particular combination of number of cores and frequency adopted. However, for sake of readability, we do not explicitly account for such a dependency in the notation.   

A first observation of the fitting curves for each particular server in Figure~\ref{fig:allCores} reveals that the power for near-zero load is almost the same in curves (e.g., for \texttt{Nemesis} this value is between 84 and 85 W). Observe that it is impossible to run an experiment in which the load of the CPU is actually zero to obtain the baseline power consumption of a server. %However, the previous observation makes us believe that the baseline can be safely assumed to be the observed common value.
However, all the fitting curves converge to a similar value for $\rho \rightarrow 0$, which can be assumed to represent the baseline power consumption.

A second observation is that for one core the curves grow linearly with the load. However, as soon as two or more cores are used, the curves are clearly concave, which implies that for a fixed frequency the efficiency grows with the load (we will discuss later the efficiency in terms of number of active cycles per energy unit). 

A third observation is that frequency does not significantly impact the power consumption when the load is low. In contrast, at high load, the consumption clearly increases with the CPU frequency. More precisely, the power consumption grows superlinearly with the frequency, for a fixed load and number of cores. This is particularly evident in the curves characterizing \texttt{Erdos}, which is the most powerful among our servers.  

    \begin{figure*}[!tb]
      \centering
      %\subfigure[survivor - Reading]{\label{fig:MAread}\includegraphics[width=0.5\textwidth, height=5cm]{./figs/reading_survivor.eps}}%
      %\subfigure[survivor - Writing]{\label{fig:MAwrite}\includegraphics[width=0.5\textwidth, height=5cm]{./figs/writing_survivor.eps}}
      \subfigure[Power consumption during reading (\texttt{Nemesis}).]{\label{fig:MBread}\includegraphics[width=0.5\textwidth, height=5cm, trim= 0 20mm 0 0, clip=true]{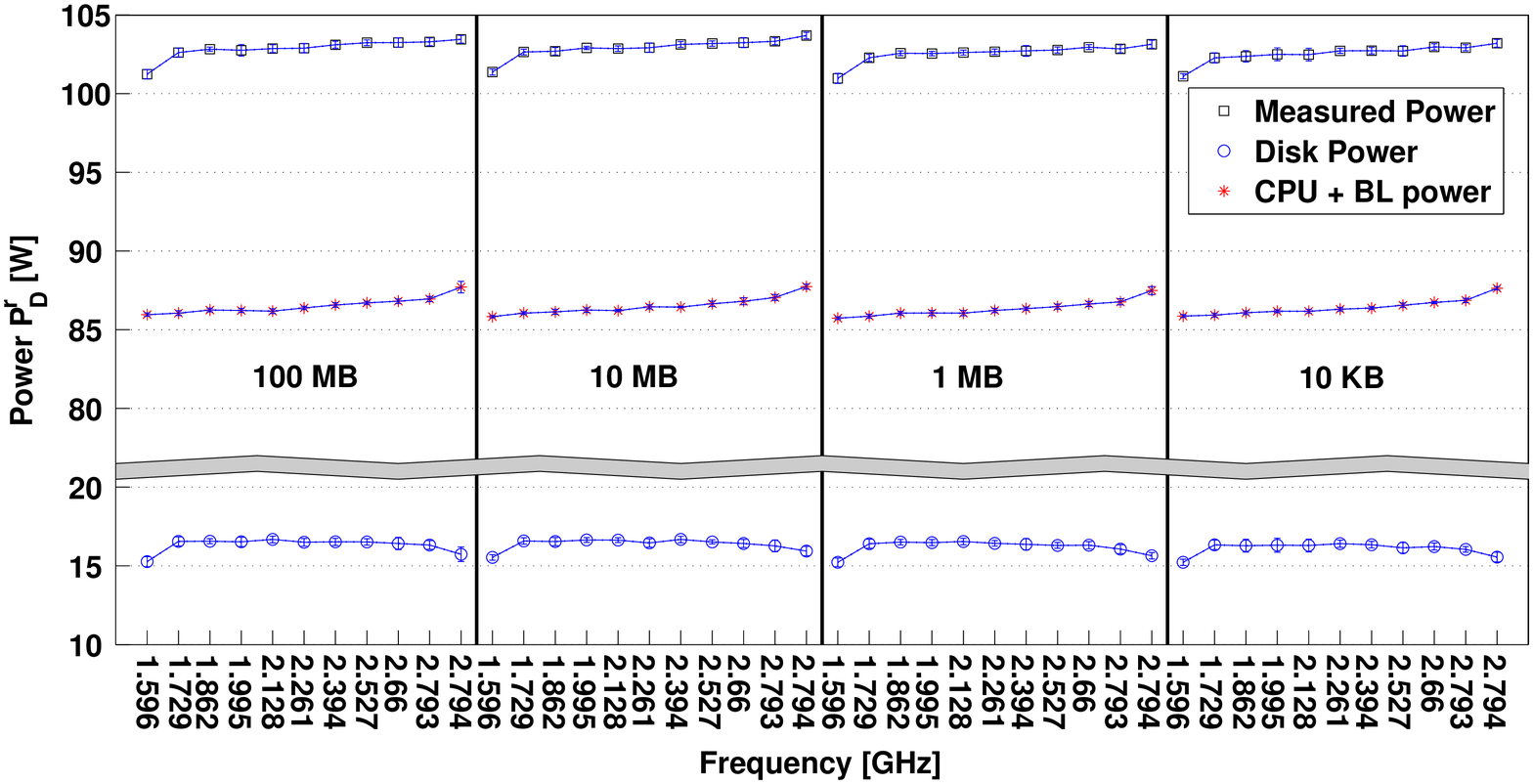}}%
      \subfigure[Power consumption during writing (\texttt{Nemesis}).]{\label{fig:MBwrite}\includegraphics[width=0.5\textwidth, height=5cm, trim= 0 20mm 0 0, clip=true]{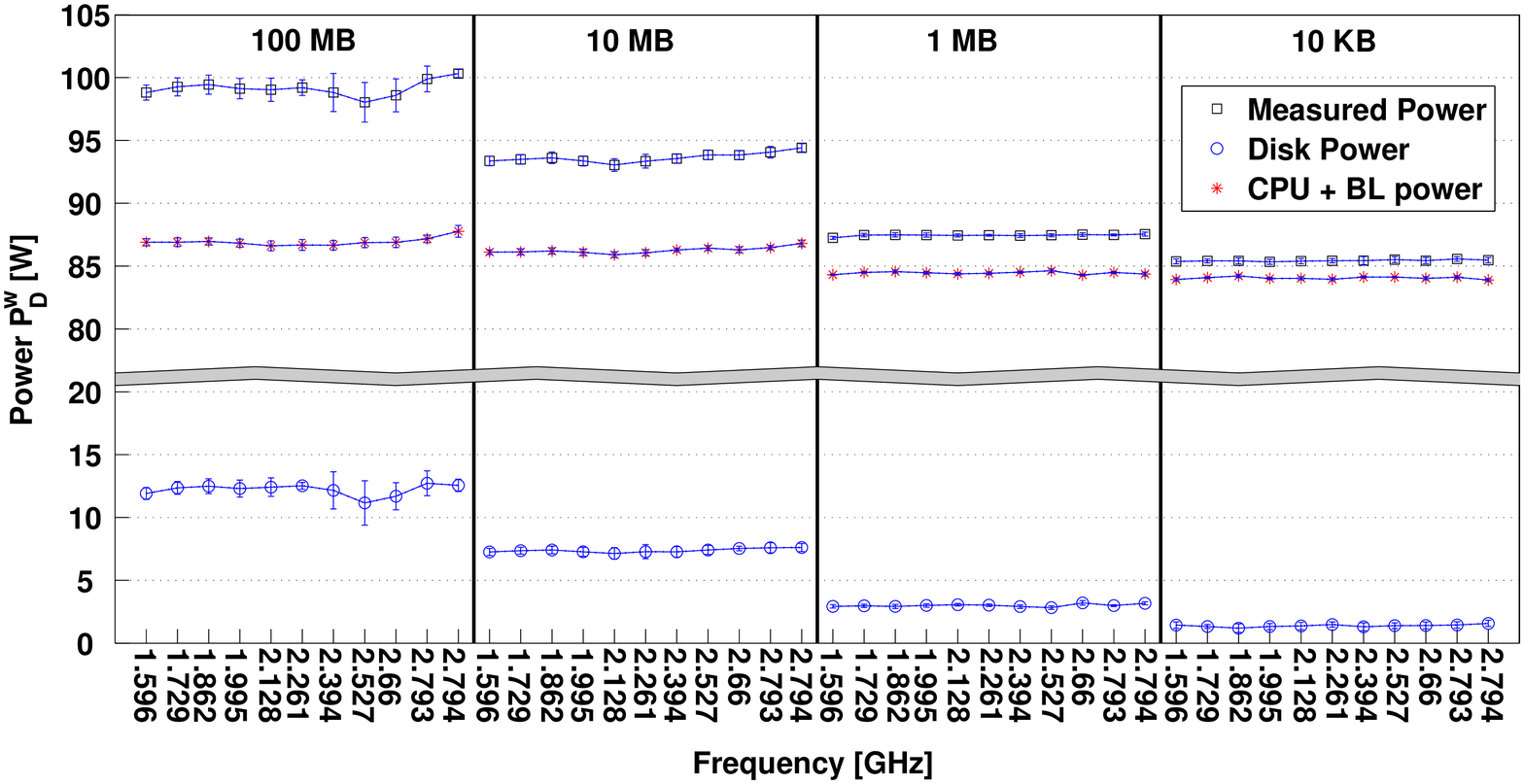}}
      \\
      \vspace{-4mm}
      \subfigure[Power consumption during reading (\texttt{Erdos}).]{\label{fig:MCread}\includegraphics[width=0.5\textwidth, height=5cm, trim= 0 10mm 0 0, clip=true]{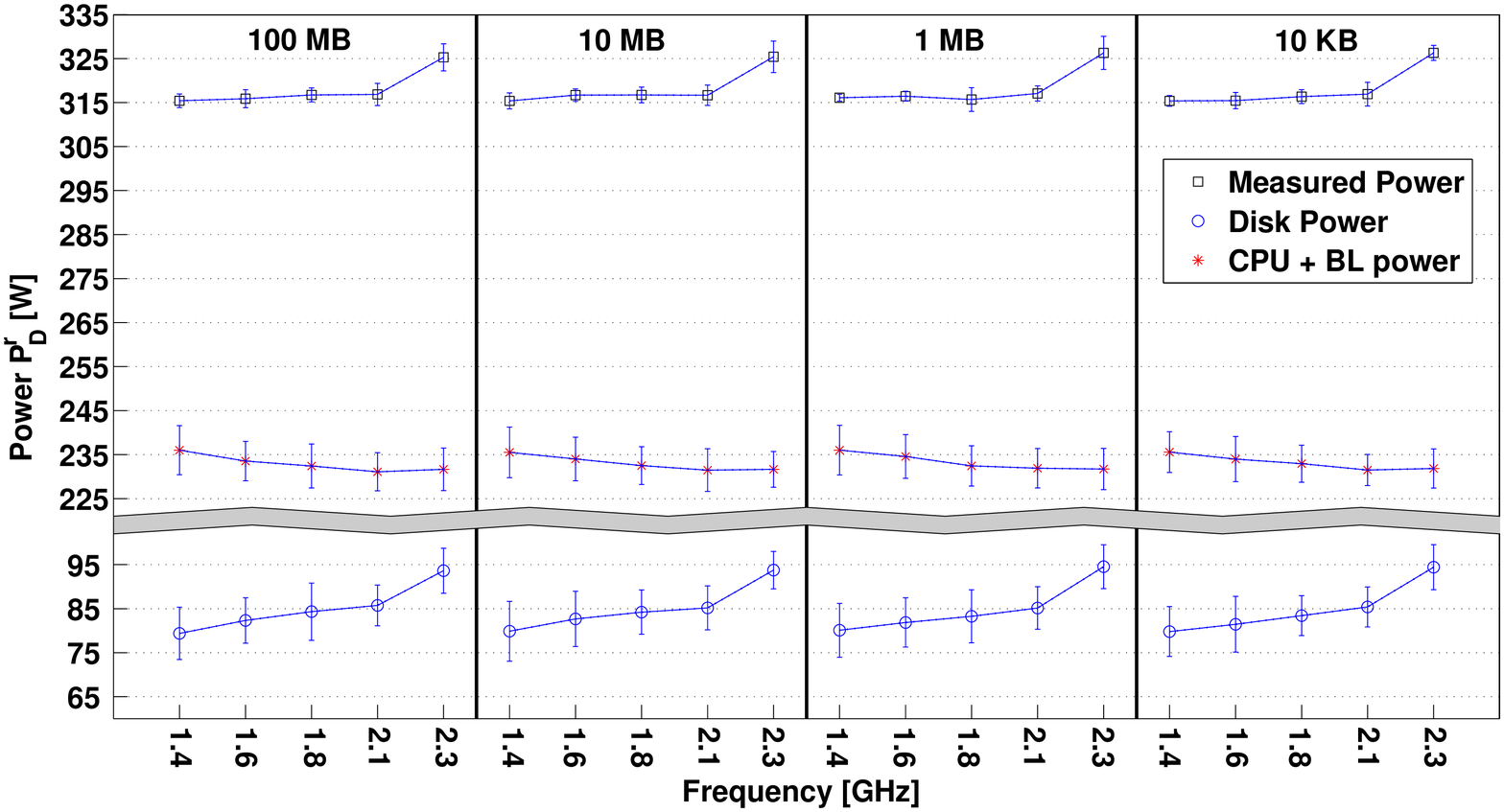}}%
      \subfigure[Power consumption during writing (\texttt{Erdos}).]{\label{fig:MCwrite}\includegraphics[width=0.5\textwidth, height=5cm,  trim= 0 10mm 0 0, clip=true]{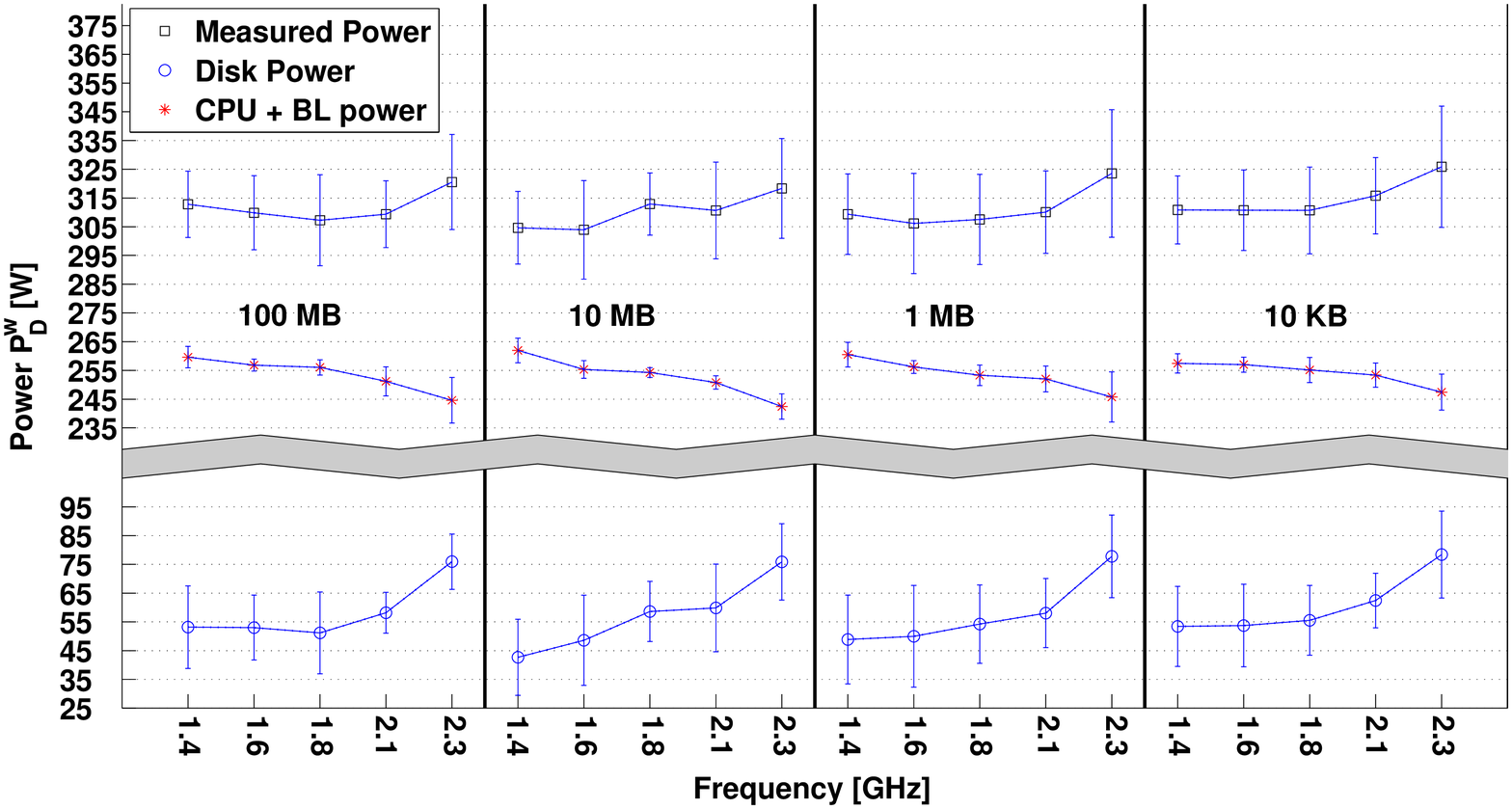}}
      \vspace{-3mm}
      \caption{Instantaneous power consumption for a reading/writing operations. Results are presented for every frequency and for $4$ different block sizes for each one of our servers.}
      \vspace{-4mm}
      \label{fig:hddinstant}
    \end{figure*}

From the previous figures it emerges that the power consumption due to CPU and baseline can be minimized by selecting the right number of active cores and a suitable CPU frequency. Similarly, we can expect that the energy efficiency, defined as number of active cycles per energy unit, can be maximized by tuning the same operational parameters. We graphically represent the impact of operation parameters on power consumption and energy efficiency in Figures~\ref{fig:nemesisenvelops} and \ref{fig:erdosenvelops} respectively for \texttt{Nemesis} and \texttt{Erdos} (results for \texttt{Survivor} are similar to the ones shown for \texttt{Nemesis} and are omitted).
In particular, Figures~\ref{fig:nemesisEnv} and \ref{fig:erdosEnv} report all possible fitting curves for the power consumption measurements, plus a curve marking the lowest achievable power consumption at a given load. We name such a curve ``minimal power curve'' $P_{\min}(\rho)$, and we observe that $(i)$ it only depends on the load $\rho$, and $(ii)$ it is a piecewise concave function, which makes it suitable to formulate power optimization problems.   
Finally, to evaluate the energy efficiency of the CPU, we report in Figures~\ref{fig:nemesisEff} and \ref{fig:erdosEff} the number of active cycles per energy unit obtained from our measurements respectively for \texttt{Nemesis} and \texttt{Erdos}. We compute the power due to active cycles as the power $P_{BC}-\alpha_0$, i.e., by subtracting the baseline consumption from $P_{BC}$, and we obtain the efficiency $\eta_C$ by dividing the load (in active cycles per second) by the power due to active cycles: 
\begin{equation}
\eta_C = \frac{\rho}{P_{BC}(\rho)-\alpha_0}.
\end{equation}
Also in this case we show the curve that maximizes the efficiency at a given load, which we name ``Maximal efficiency curve'' $\eta_{\max}(\rho)$. Interestingly, we observe that $(i)$ $\eta_{\max}(\rho)$ presents multiple local maxima, $(ii)$ for a given configuration of frequency and number of active cores, the efficiency is maximized at the highest achievable load, $(iii)$ all local maxima corresponds to the use of all available active cores, but $(iv)$ the absolute maximum is {\it not} achieved neither at the highest CPU frequency nor at the lowest.

\subsection{Disks}
\label{sub:meas:disks}

%{\color{cyan}
%
%- figures: avg+stddev and curves for $P_{tot}$, $P_{BC}$, and $P_D$ computed as difference of the previous two. +figures for efficiency
%
%- values are relatively small but not negligible w.r.t. the baseline.
%
%- writing power depends on block size $B$
%
%- reading power: almost constant for Nemesis/Survivor, but grows with the frequency for Erdos (different disks and file systems)
%
%- efficiency of reading is constant. efficiency for writing changes with $B$, not with $f$ (saturation observed: due to seeking time --> concavity)
%
%}

We now characterize the power and energy consumption of disk I/O operations. During the experiments, we continuously commit either read or write operations, while keeping the CPU load $\rho$ as low as possible (i.e., we disconnect the network and we do not run other tasks). Still, the power measurements obtained during the disk experiments contain both the power used by the disk and power due to CPU and baseline. Indeed, Figure~\ref{fig:hddinstant} shows, for each experiment, the total measured power $P_{t}$, the power $P_{BC}$ computed according to Eq.~\ref{eq:PBC} at the load $\rho$ measured during the experiment, and the power due to disk operations, computed as:
\begin{equation}
P_D^x = P_{t} - P_{BC}(\rho), \quad x \in \{r,w\}, 
\end{equation}   
where superscripts $r$ and $w$ refer to reading and writing operations, respectively.
We test sequentially all the available frequencies for each server (see Table~\ref{tab:servers}), and I/O block sizes ranging from $10$ {\it KB} to $100$ {\it MB}.
Figure~\ref{fig:hddinstant} shows average and standard deviation of the measures over $10$ experiment repetitions. Results for \texttt{Survivor} are omitted since they are like \texttt{Nemesis}' results. Indeed, \texttt{Survivor} and \texttt{Nemesis} have similar disks and file systems, while \texttt{Erdos} is equipped with SAS disks with RAID. 
In all cases shown in the figure, the disk power is small but not negligible with respect to the baseline consumption. Furthermore, we can observe that the two servers presented behave differently. Indeed, while the power consumption due to writing is affected both by the block size $B$ for both machines, we observe that \texttt{Nemesis}' disk writing power $P_D^w$ is not affected by the CPU frequency, while \texttt{Erdos}' results show an increase with the frequency. Moreover, the results obtained with \texttt{Erdos} are affected by a substantial amount of variability in the measurements, which we believe is due to the caching operations enforced by the RAID mechanism in \texttt{Erdos}. 

\begin{figure}[!tb]
      \centering
       	%\vspace{-3mm}
      \includegraphics[width=0.52\textwidth, trim=25mm 0 0 10mm , clip=true]{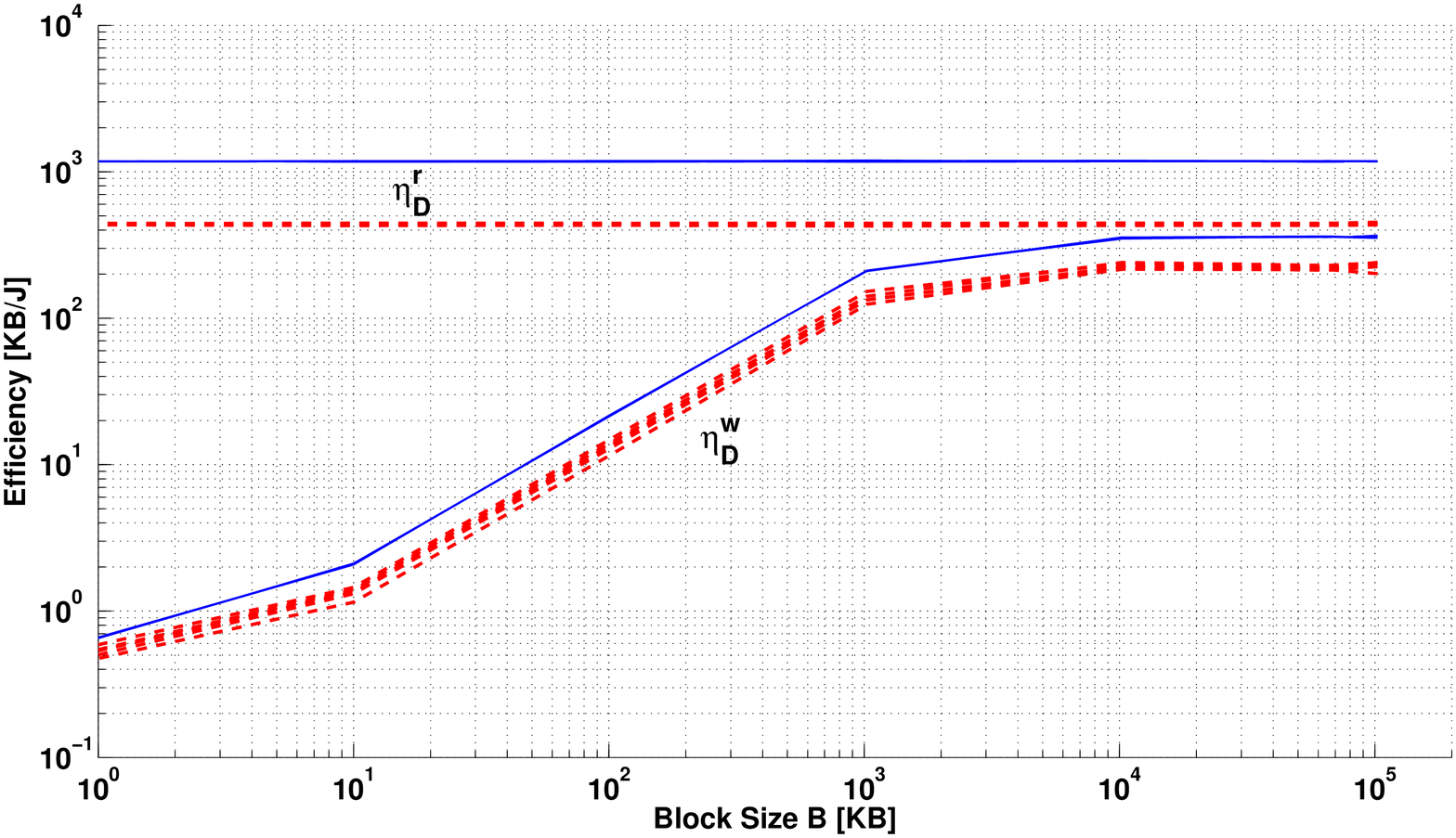}\\
       	\vspace{-4mm}
      \caption{Disk reading and writing efficiencies for \texttt{Erdos} (red dotted lines) and \texttt{Nemesis} (blue solid lines).}
       	\vspace{-4mm}
      \label{fig:diskEfficiency}
\end{figure}

Similarly to what was described for the CPU, we now comment on the energy efficiencies $\eta_D^r$ and $\eta_D^w$ of disk reading and writing operations. 
Figure~\ref{fig:diskEfficiency} reports efficiency as a function of the I/O block size, and shows one line per each CPU frequency. The efficiency is computed by subtracting the baseline power from the total power, and by measuring the volume $V$ of data read or written in an interval $T$: 
\begin{equation}
\eta_D^x = \frac{V}{P_D^x T}, \quad x \in \{r,w\}.
\end{equation}
We can observe that results are similar for all the servers. Specifically, the efficiency of reading is almost constant at any frequency and for each block size, while writing is more efficient with large block sizes. We also observe that the efficiency changes very little with the adopted CPU frequency. Another observation is that the efficiency saturates to a disk-dependent asymptotic value, which is due to the mechanical constraints of the disk (e.g., due to the non-negligible {\it seek} time, the number of read/write operations per second is limited). In addition, although not visible in the figure due to the log-scale adopted, $\eta_D^w$ is a concave function of the block size $B$.

%By observing the instantaneous consumption figures we notice that, when reading, although it exists, the influence of the frequency in the consumption is very small, being the value almost constant. On the other hand, when we are writing

%\subsection{Memory}
%\label{sub:meas:memory}

\subsection{Network}
\label{sub:meas:network}

The last server component that we characterize via measurements is the network card. Similarly to the cases described previously, we run experiments in which only the 
operating system and our test scripts are active. In this case, we run a script to either transmit or receive UDP packets over a gigabit Ethernet connection and count the system 
active cycles $\rho$. We measure the total power consumption $P_t$ during the experiment, so that the power due to network activity can be then estimated as follows: 
\begin{equation}
P_N^x = P_{t} - P_{BC}(\rho), \quad x \in \{s,r\}, 
\end{equation}     
where superscripts $s$ and $r$ refer to the sender and the receiver cases, respectively.

In the experiments, we sequentially test all the available frequencies for each server (see Table~\ref{tab:servers}), and fix the packet size $S$ and the UDP transmission rate within the achievable set of rates (which depends on the packet size, e.g., $< 950$ {\it Mbps} for $1470$-B packets).  
We report results for the network energy in terms of efficiencies $\eta_N^s$ and $\eta_N^r$ (volume of data transferred per unit of energy). These efficiencies are  
computed as follows: 
\begin{equation}
\eta_N^x = \frac{R}{P_N^x}, \quad x \in \{s, r\}, 
\end{equation}
where $R$ is the transmission rate during the experiment.  

\begin{table*}[!ht]
\caption{Polynomial fitting for network efficiency. The table reports empirically evaluated coefficients for Eq.~\ref{eq:effN} (coefficients $\beta_1$ are expressed in $W^{-1}$ while coefficients $\beta_2$ are in $W^{-1}\cdot \textit{bps}^{-1}$).}
\vspace{-3mm}
\label{tab:calibrationNetwork}
\scriptsize
\begin{tabular}{||c||c|c|c|c|c||c|c|c|c|c||}
\hline
%\# of cores & \multicolumn{4}{||c||}{coefficients} & \# of cores & \multicolumn{4}{||c||}{coefficients} & \# of cores & \multicolumn{4}{||c||}{coefficients} \\ 
%\hline
& \multicolumn{10}{c||}{\bf RECEIVER} \\
\hline
&  \multicolumn{5} {c||}{\texttt{Survivor}} & \multicolumn{5} {c||}{\texttt{Nemesis}}\\
\hline
&  \backslashbox{\kern-0.5em pck size\kern-0.5em}{\kern-1em freq} & $1.2$ {\it GHz} & $1.6$  {\it GHz} & $1.867$  {\it GHz} & $2.133$  {\it GHz} &\backslashbox{\kern-0.5em pck size\kern-0.5em}{\kern-2em freq \kern-1em} & $1.596$  {\it GHz} & $1.995$   {\it GHz}& $2.394$ {\it GHz} & $2.794$  {\it GHz}\\
\hline 
$\beta_1$  &  \multirow{2}{*}{64 {\it B}}	&  1.751e-2 & 1.314e-2 & 1.268e-2 & 1.254e-2 & \multirow{2}{*}{64 {\it B}}  &  1.491e-2 & 1.410e-2 & 1.330e-2 & 1.227e-2 \\
%5.46377e-2 & 4.4134e-2 & 3.5513e-2 & 4.37822e-2  \\
$\beta_2]$ &				& 1.904e-5	& 2.160e-5 & 1.395e-5 & 1.031e-5 & 				& &&&\\
%-0.0000175146 	& 0.0000207872 	& 0.0000455446 & $-5.54432*10^{-6}$\\
\hline
$\beta_1$ &	\multirow{2}{*}{500 {\it B}}			& 1.736e-2 & 1.386e-2 & 1.144e-2 & 9.962e-3 & 	\multirow{2}{*}{500 {\it B}} & 	1.565e-2 	& 1.234e-2 & 1.107e-2 & 1.074e-2  \\
$\beta_2$ & 				&  	2.627e-6		& 	1.595e-6		& 2.836e-6 &	3.541e-6 &		& 	&&&\\	
%2.88449e-5		& 0.0000214837 		& 0.0000244173 &0.0000297506\\
\hline 
$\beta_1$ &  \multirow{2}{*}{1000 {\it B}}	& 1.560e-2 & 1.296e-2 & 1.132e-2 & 1.029e-2 & \multirow{2}{*}{1000 {\it B}}  &  1.170e-2	& 9.451e-3 & 7.712e-3  & 7.448e-3 \\
$\beta_2$ & 				& 3.155e-6 & 1.736e-6 & 1.080e-6 & 	1.208e-6		&	& &&&\\
%7.94555e-6 	& 0.0000129492 	& $-2.00543*10^{-6}$ & $-1.45133*10^{-6}$\\
\hline
$\beta_1$ & \multirow{2}{*}{1470 {\it B}}		&  1.497e-2  & 1.216e-2  & 1.073e-2  & 2.684e-2 & 	\multirow{2}{*}{1470 {\it B}}	& 1.072e-2 & 8.849e-3 & 8.207e-3  & 8.040e-3 \\
$\beta_2$ & 				&  	3.231e-6		& 	4.006e-6		& 3.533e-6 & 	-4.746e-6 &		& 	&&&\\		
%$-2.33781*10^{-6}$	& 	$-1.09768*10^{-7}$	& $-1.42442*10^{-7}$ & $-2.31226*10^{-6}$\\
\hline
\hline
& \multicolumn{10} {c||}{\bf SENDER} \\
\hline
&  \multicolumn{5} {c||}{\texttt{Survivor}} & \multicolumn{5} {c||}{\texttt{Nemesis}}\\
\hline
&  \backslashbox{\kern-0.5em pck size\kern-0.5em}{\kern-1em freq} & $1.2$ {\it GHz} & $1.6$  {\it GHz} & $1.867$  {\it GHz} & $2.133$  {\it GHz} &\backslashbox{\kern-0.5em pck size\kern-0.5em}{\kern-2em freq \kern-1em} & $1.596$  {\it GHz} & $1.995$   {\it GHz}& $2.394$ {\it GHz} & $2.794$  {\it GHz}\\
\hline 
%&  \multicolumn{5} {c||}{Survivor} & \multicolumn{5} {c||}{Nemesis}\\
%\hline
%\# of cores & \multicolumn{4}{||c||}{coefficients} & \# of cores & \multicolumn{4}{||c||}{coefficients} & \# of cores & \multicolumn{4}{||c||}{coefficients} \\ 
%\hline
%&  \backslashbox{\kern-0.5em pck size\kern-0.5em}{\kern-1em freq} & $1.2~GHz$ & $1.6~GHz$ & $1.867~GHz$ & $2.133~GHz$ &\backslashbox{\kern-0.5em pck size \kern-0.5em}{\kern-2em freq \kern-1em} & $1.596~GHz$ & $1.995~GHz$ & $2.394~GHz$ & $2.794~GHz$\\
%\hline 
$\beta_1$ & 64 {\it B}& 2.239e-2 & 1.802e-2 & 1.582e-2	& 1.462e-2 & 64 {\it B}  & 1.642e-2 & 1.313e-2 & 1.029e-2 & 8.625e-3 \\
%$\alpha_0$ &				& 3.795e-29	& 3.795e-29	& 3.795e-29	& & 				& 3.795e-29 	& 3.795e-29 	& 3.795e-29 & \\
\hline
$\beta_1$ & 500 {\it B}	& 1.742e-2 & 1.576e-2	& 1.429e-2 & 2.205e-2 & 	500	{\it B}		& 1.599e-2	& 1.130e-2 & 1.234e-2 & 1.014e-2 \\
%$\alpha_0$ & 				&  			& 			& &			& 				& 			& 			& &\\
\hline 
$\beta_1$ & 1000 {\it B}	& 1.784e-2 & 1.634e-2 & 1.454e-2 & 2.230e-2 & 1000 {\it B}  & 1.767e-2 & 1.781e-2 & 1.824e-2 & 1.179e-2\\
%$\alpha_0$ & 				& 3.795e-29	& 3.795e-29	& 3.795e-29	& & 				& 3.795e-29 	& 3.795e-29 	& 3.795e-29 &\\
\hline
$\beta_1$ & 1470 {\it B}	& 1.801e-2 & 1.620e-2 & 1.461e-2 & 2.369e-2 & 	1470 	{\it B}		& 1.703e-2	& 1.863e-2 & 1.279e-2 & 1.134e-2\\
%$\alpha_0$ & 				&  			& 			&& 			& 				& 			& 			& &\\
\hline
\end{tabular}
	\vspace{-5mm}
\end{table*}

\begin{figure*}[ht]
	\centering
%	\hspace{-0.7in}
	\subfigure[Receiver network efficiency (\textbf{Survivor}).]{
		\includegraphics[angle=-90,width=0.43\linewidth]{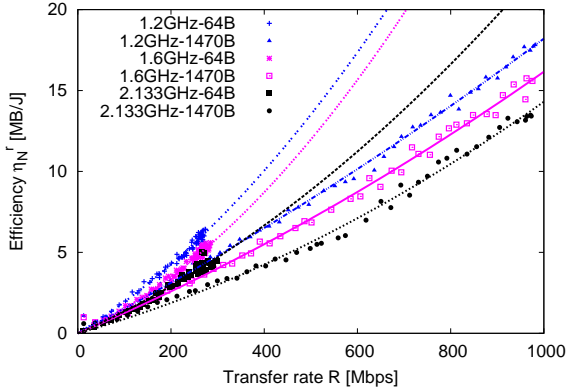}
		\label{fig:survivorServerEffFreq}
		}
	\quad
	\subfigure[Receiver network efficiency (\textbf{Nemesis}).]{
		\includegraphics[angle=-90,width=0.43\linewidth]{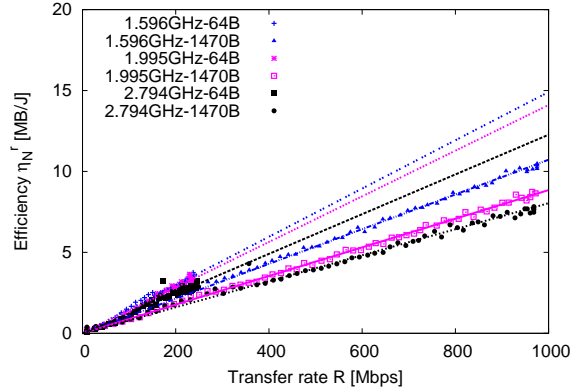}
		\label{fig:eeeServerEffFreq}
	}
	\\
	\vspace{-3mm}
	\subfigure[Sender network efficiency (\textbf{Survivor}).]{
		\includegraphics[angle=-90,width=0.43\linewidth]{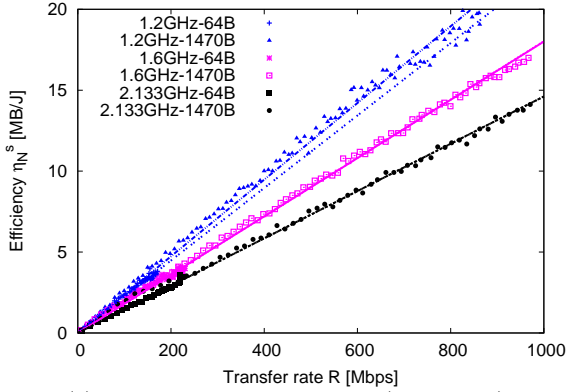}
		\label{fig:survivorClientEffFreq}
		}
	\quad
	\subfigure[Sender network efficiency (\textbf{Nemesis}).]{
		\includegraphics[angle=-90,width=0.43\linewidth]{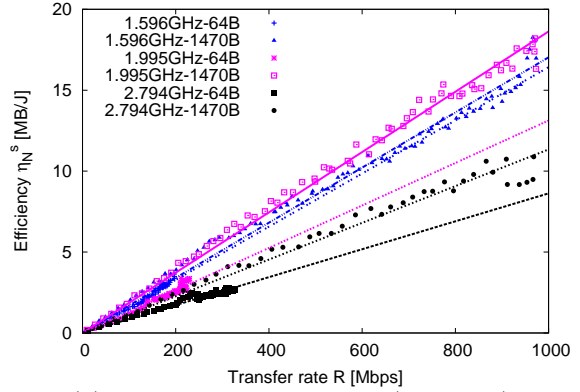}
		\label{fig:eeeClientEffFreq}
	}\\
		\vspace{-4mm}
	\caption{Network efficiencies for different frequencies and $64$-{\it B} and $1470$-{\it B} packets.}
	\label{fig:networkEfficiency}
		\vspace{-3mm}
\end{figure*}

Figure~\ref{fig:networkEfficiency} shows the network efficiencies of \texttt{Nemesis} and \texttt{Survivor} averaged over $3$ samples 
per transmission rate $R$.\footnote{Network results are obtained by using a point-to-point Ethernet connection between two controlled servers. 
Since \texttt{Erdos} is located in a different building with respect to \texttt{Nemesis} and \texttt{Survivor}, it was not possible to test the network efficiency of \texttt{Erdos}.}
For sake of readability, the figure only shows results for the extreme value used for the packet size, and for three CPU frequencies: the lowest, the highest, and an intermediate 
frequency in the set of available frequencies reported in Table~\ref{tab:servers} for \texttt{Nemesis} and \texttt{Survivor}. The figure also reports the polynomial fitting curves for efficiency, which we found to be at most of second order. Since the efficiency is represented in terms of network activity only, in the fitting we force the zero-order coefficient of the polynomials to be $0$. Therefore, we can use the following expression to characterize the network efficiencies of our servers: 
\begin{equation}
\label{eq:effN}
\eta_N^x = \beta_1 R + \beta_2 R^2, \quad x \in \{s,r\}, 
\end{equation}  
where the $\beta_i$ coefficients are computed by minimizing the least square error of the fitting.
Table~\ref{tab:calibrationNetwork} gives the fitting coefficients for sending and receiving efficiencies for the cases shown in   Figure~\ref{fig:networkEfficiency} and for other tested configurations. 

From both the figure and the table, we can observe that efficiencies are almost linear or slightly superlinear with the transfer rate, e.g., the receiving efficiency of \texttt{Survivor} exhibits 
an evident quadratic behavior. Indeed, our measurements show that the network power consumption is independent from the throughput, which is a well known result for legacy Ethernet devices. In fact, the NICs of our servers are not equipped with power saving features like, e.g., the recently standardized IEEE 802.3az~\cite{802.3az}.

In all cases, the efficiency is strongly affected by the selected CPU frequency. Moreover, efficiency is also affected by packet size, although the impact of packet size changes from server to server, e.g., \texttt{Survivor} sending efficiency is only slightly affected by it.  

\begin{figure}[t!]
\centering
%\vspace{-2mm}
    \includegraphics[angle=-90,width=0.9\linewidth]{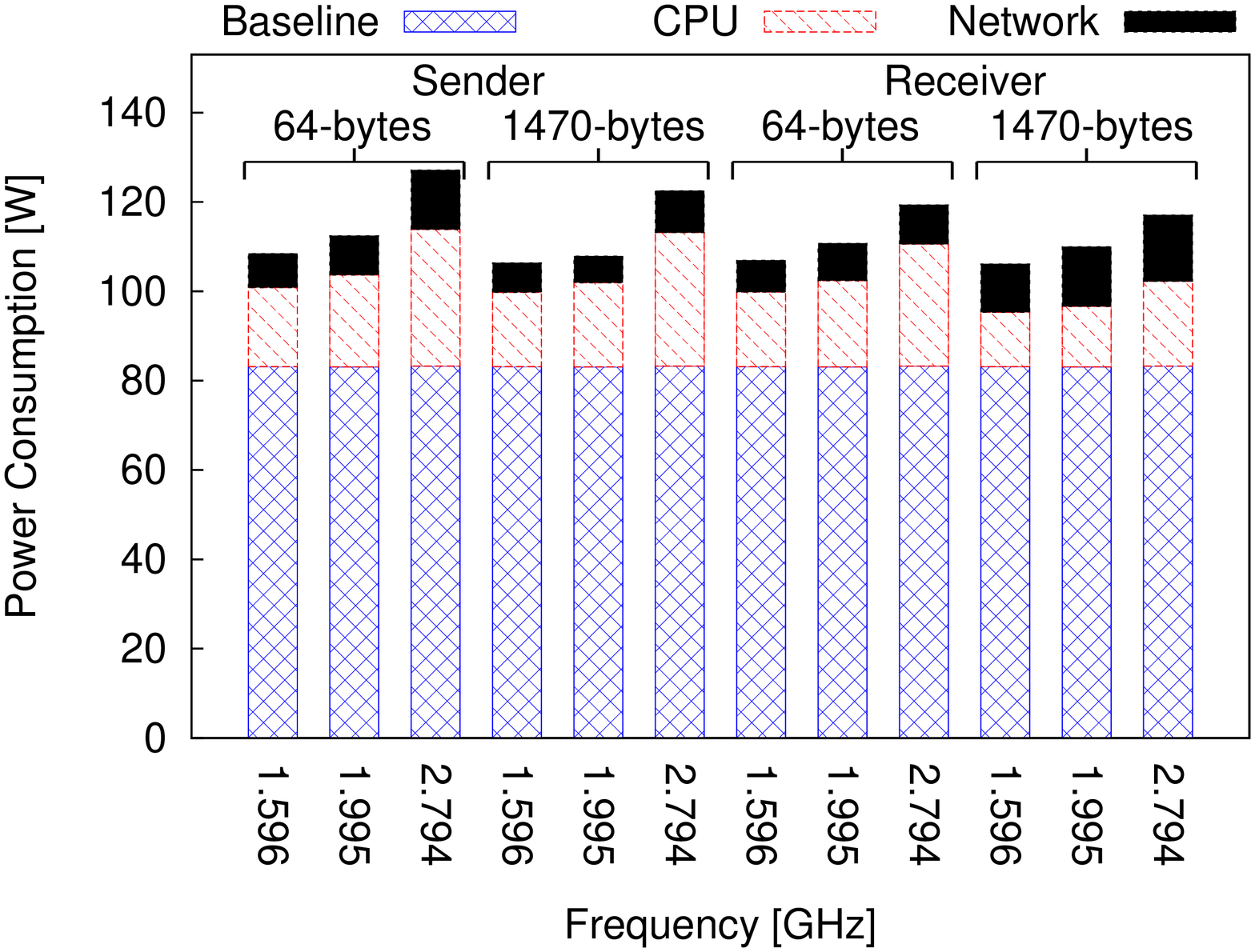}
    	\vspace{-3mm}
    \caption{Power consumption with network activity for \texttt{Nemesis} ($64$-{\it B} experiments were run with a  transmission rate $R=150$ {\it Mbps}, while 
    $R=400$ {\it Mbps} for the experiments with $1470$-{\it B} packets).}
    	\vspace{-5mm}
    \label{fig:NetworkHistogram}
\end{figure}

Another observation is that, depending on the packet size and frequency used, sending can be more energy efficient than receiving at a given transmission rate, and using the highest CPU frequency is never the most efficient solution. Note also that the efficiency decreases with the packet size, although this effect is particularly evident at the receiver side, while it only slightly impacts the efficiency of the packet sender. However, network activity also causes non-negligible CPU activity, as shown in Figure~\ref{fig:NetworkHistogram} for a few experiment configurations for \texttt{Nemesis}.  Overall, the lowest CPU frequency yields the lowest total power consumption during network activity periods.

\section{Estimating Energy Consumption}
\label{sec:model}

While the results presented in the previous sections are useful to understand the power consumption pattern of CPU, disk and network, we believe that a much more important use of these results is to estimate the energy consumption of applications. In this section we describe how this could be done from simple data about the application, and validate the proposed process by estimating the energy consumed by map-reduce Hadoop computations.

\subsection{Energy Estimation Hypothesis}

The process we propose to estimate the energy consumed $E_{app}$ by an application has as basic assumption that this energy is essentially the sum of the baseline energy $E_B$ (the baseline power times the duration of the execution), the energy consumed by the CPU $E_C$, the energy consumed by the disk $E_D$, and the energy consumed by the network interface $E_N$. I.e.,
\begin{equation}
\label{eq:estimate}
E_{app} = E_B + E_C + E_D + E_N.
\end{equation}
Hence, the process of estimating $E_{app}$ is reduced to estimating these four terms. In order to estimate the first two terms, we need to know the total number of active cycles that the application will execute, $C_{app}$, and the load $\rho_{app}$ (in ACPS) that the execution will incur in the CPU. From this, the total running time $T_{app}$ can be computed as
\begin{equation}
T_{app}=C_{app}/\rho_{app}.
\end{equation}
Then, once the number of cores and the frequency that will be used have been defined, it is also possible to estimate the baseline power plus CPU power, $P_{BC}$, from the fitting curves of Figure~\ref{fig:allCores}. This allows to estimate the sum of the first two terms of Eq.~\ref{eq:estimate} as
\begin{equation}
\label{eq:estimateBC}
E_B + E_C = P_{BC} T_{app} = P_{BC} C_{app}/\rho_{app}.
\end{equation}

The energy consumed by the disk is simply the energy consumed while reading and writing, i.e., $E_D = E_D^r + E_D^w$. To estimate these latter values, the block size to be used has to be decided, from which we can obtain an estimate of the efficiency of reading, $\eta_D^r$, and writing, $\eta_D^w$ (see Figure~\ref{fig:diskEfficiency}). These, combined with the total volume of data read and written by the application, denoted as $V_D^r$ and $V_D^w$ respectively, allow to obtain the estimate energy as
\begin{equation}
\label{eq:estimateD}
E_D = \frac{V_D^r}{\eta_D^r} + \frac{V_D^w}{\eta_D^w}.
\end{equation}

Finally, to estimate $E_N$, the transfer rate $R$ and packet size $S$ has to be chosen, which combined with the frequency used, yield sending and receiving efficiencies $\eta_N^s$ and $\eta_N^r$ (see Figure~\ref{fig:networkEfficiency}).
Then, if the total volume of data to be sent and received is $V_N^s$ and $V_N^r$, respectively,
\begin{equation}
\label{eq:estimateN}
E_N= \frac{V_N^s}{\eta_N^s} + \frac{V_N^r}{\eta_N^r}.
\end{equation}
All is left to do to obtain the estimate $E_{app}$ is to add up the values obtained in Equations \ref{eq:estimateBC}, \ref{eq:estimateD}, and \ref{eq:estimateN}.
%\begin{comment}
%  \begin{figure*}[!tb]
%      \centering
%      \subfigure[Sender 1470 bytes]{\label{fig:hadSend1470}	\includegraphics[width=\columnwidth]{./figs/hadoop/hadSend1470.eps}}
%      \subfigure[Sender 64 bytes]{\label{fig:hadSend64}	\includegraphics[width=\columnwidth, trim= 24mm 6mm 21mm 10mm, clip=true]{./figs/hadoop/hadSend64.eps}}
%      \subfigure[Receiver 1470 bytes]{\label{fig:hadRec1470}	\includegraphics[width=\columnwidth, trim= 24mm 6mm 21mm 10mm, clip=true]{./figs/hadoop/hadRec1470.eps}}
%      \subfigure[Receiver 64 bytes]{\label{fig:hadRec64}	\includegraphics[width=\columnwidth, trim= 24mm 6mm 21mm 10mm, clip=true]{./figs/hadoop/hadRec64.eps}}
%     \caption{Hadoop results}
%     \label{fig:allCores}
%   \end{figure*}
%\end{comment}

\begin{figure}[tb!]
\vspace{-0mm}
	\centering
	\subfigure[Sender side.]{
		\includegraphics[width=0.95\columnwidth, trim= 0 5mm 0 5mm, clip=true]{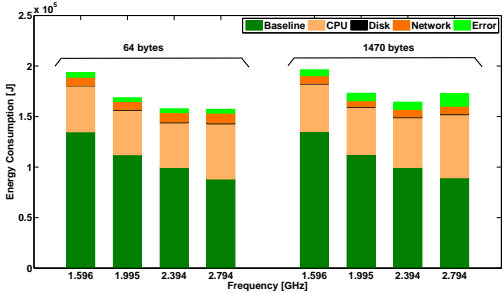}
		\label{fig:hadoopSender}
	}\\
	\vspace{-3mm}
	\subfigure[Receiver side.]{
		\includegraphics[width=0.95\columnwidth, trim= 0 5mm 0 5mm, clip=true]{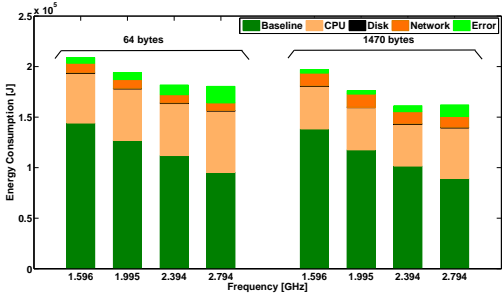}
		\label{fig:hadoopReceiver}
	}\\
	\vspace{-3mm}
	\caption{Comparison of the real versus estimated energy consumption values.}
	\vspace{-5mm}
	\label{fig:hadoop}
\end{figure} 

\subsection{Empirical Validation}

We test now the process and hypothesis presented above for the estimation of the energy consumed by an application. For that, we have chosen to execute in \texttt{Nemesis} a map-reduce Hadoop application that computes several iterations of the pagerank algorithm on an Erdos-Renyi random (directed) graph with 1 million nodes and average degree 5. Since the pagerank application does not use the network, while it is running we execute another process generating network traffic.
This provides a richer experiment.
 
An execution of the pagerank application has three phases: preprocessing, map-reduce, and postprocessing. On its side, the map-reduce phase is a sequence of several homogeneous iterations of the pagerank algorithm. For simplicity, we only estimate the energy consumed during the map-reduce phase of the pagerank algorithm. In our experiments we run in \texttt{Nemesis} one instance of the pagerank application with $10$ iterations in its map-reduce phase for each one of the $11$ available frequencies. We run 
this experiment $4$ times, each with different characteristics of the network traffic generated in parallel. In particular, we run experiments with \texttt{Nemesis} behaving as a sender and as a receiving, and using packets of 64 and 1470 bytes.
Instead of estimating the energy for the whole sequence of $10$ iterations, it is simpler to estimate the energy for every iteration separately.
Then, for each iteration $i$ we can register the total active cycles executed $C_{app}^i$, the time consumed $T_{app}^i$, and the volume of data read and written, 
$V_D^{r,i}$ and $V_D^{w,i}$, respectively, and the transfer rate $R$ (the same for all iterations: $150$ {\it Mbps} for experiments with $64$-{\it B} packets and 
$400$ {\it Mbps} for experiments with $1470$-{\it B} packets).

Unfortunately, we cannot measure the instantaneous CPU load. Instead, we assume that the CPU load is the same during the execution for a given frequency and network configuration. Hence we estimate it as $\rho_{app}^i=C_{app}^i / T_{app}^i$. Then, from this value we obtain the estimate of the instantaneous power $P_{BC}^i$ using the fitting curves as described above. Finally, using Eq.~\ref{eq:estimateBC} we compute the estimate $E_B^i + E_C^i$.

In order to estimate the energy consumed by the disk operations, we use the fact that Hadoop uses a block size of $64$ {\it MB}. This allows us to estimate the reading and writing efficiencies, $\eta_D^{r,i}$ and $\eta_D^{w,i}$ (see Figure~\ref{fig:diskEfficiency}).
%(see Figure~\ref{fig:nemesisDiskEff}). 
Combining these values with the measured volume of data read and written ($V_D^{r,i}$ and $V_D^{w,i}$) as described in Eq.~\ref{eq:estimateD}, we obtain $E_D^i$.

Finally, to estimate the network consumption in one iteration with \texttt{Nemesis} sending traffic (resp., receiving traffic), the sending efficiency $\eta_N^s,$ (resp., receiving efficiency $\eta_N^r$) is obtained from the transfer rate $R$, and the frequency and packet size used (see Figure~\ref{fig:networkEfficiency}). The amount of data sent (reps., received) is obtained as the product of the rate $R$ and the time $T_{app}^i$. Then, the energy of the network is obtained using Eq.~\ref{eq:estimateN}.

%According to these assumptions we will, first, obtain the average active cycles per second consumed during the iteration for a particular frequency and with it obtain $P_{BC}$ and, multiplying it per the iteration time, obtain the total energy due to the CPU and the baseline. In order to estimate the power consumption due to the read and write operations we first transform $\eta_D^{r,w}$, for a blocksize of $64$MB and the current frequency, from $KB/J$ to accesses per Joule (we can obtain this value by dividing the amount of read or write accesses performed per second during our read(write) experiments and dividing them by $P_D^{r,w}$). Then, by dividing the total number of write, or read respectively, accesses per the efficiency we will obtain the total energy due to the disk. Finally, we obtain the total energy consumed by the network by multiplying the time needed for each iteration times the corresponding network efficiency.

Once we have computed the energy due to the different components in iteration $i$, the total energy $E_{app}^i$ is obtained by adding them. Adding these values
for the $10$ iterations of an experiment we obtain the estimate $E_{app}$.
The (approximate) total \textit{real} energy $\hat{E}_{app}^i$ consumed by iteration $i$ is computed by obtaining the average value of the power samples we registered with our power analyzer during the iteration, and multiplying it by $T_{app}$. Again, the total energy consumed by the experiment are obtained as 
$\hat{E}_{app}=\sum_{i=1}^{10}  \hat{E}_{app}^i$.
The estimation error for each experiment is then computed as $\hat{E}_{app} - E_{app}$.

We show the results obtained for four selected frequencies (the results for the rest are similar) in Figure \ref{fig:hadoopSender}, for the sender cases, and Figure \ref{fig:hadoopReceiver} for the receiving cases. Each figure includes the results for the two packet sizes used.
As can be seen, the error is very small (always below $7\%$ of the total energy), being a bit more relevant in the case of the highest frequency.

\section{Discussion}
\label{sec:discussion}

We discuss now some of the implications of our results.
%\paragraph{Consolidation:}
We start with consolidation. It has been typically assumed that the best way of doing consolidation is to fill servers as much as possible, to reduce the total number of servers being used, hence proposing bin-packing based solutions~\cite{beloglazov2012energy, mishra2011theory, wangconsolidating} and not necessarily having frequency into account. However, the results presented in Figures~\ref{fig:nemesisEff} and \ref{fig:erdosEff} show that the highest frequency is not the most efficient one, and this has been found to be true for two different architectures (Intel and AMD). This implies that, by running servers at the optimal amount of load, and the right frequency, a considerable amount of energy could be saved.

%\paragraph{Baseline:}
A second relevant aspect is the baseline consumption of servers. The results presented for all $3$ servers show that their baselines are within a $30$-$50\%$ of the maximum consumption. Then, it is straightforward that more effort is to be done for reducing baseline consumption. For instance, a solution could consist in switching off cores in real time, not just disabling them,  or in introducing very fast transitions between active and lower energy states, i.e., to achieve real {\it suspension} in idle state. 
%, were faster so its use became a reality in data centers, as the usual solution is not suspending machines but letting them run idle.

%\paragraph{CPU load associated to Network:}
Finally, we refer to the CPU load associated to disk and network activity.
% to sending or receiving data through the network. 
It can be observed in Figure~\ref{fig:hddinstant} that disks do not incur much CPU overhead. In fact, the power consumed by CPU plus baseline does not change much across the experiments. Instead, the energy consumed by CPU due to network operations is even larger than the energy consumed by the NIC (see Figure~\ref{fig:NetworkHistogram}). Some works \cite{garcia2012energy} have already pointed out that the way the packets are handled by the protocol stack is not energy efficient. Our results reinforce this feeling and point out 
that building a more efficient protocol stack would certainly reduce the amount of energy consumed due to the network.
%!TEX root = ./eenergy2014.tex
\section{Related Work}
\label{sec:related}

There is a large body of work in the field of modeling server power consumption and its components, both theoretical and empirical. 
%One of the more common discussions has been whether 
The consumption of servers 
%follow a linear or not linear behavior depending on the load or utilization of the machine, in fact, we can find theoretical works 
has been assumed as linear
% models for the consumption, as 
e.g., by Wang {\it et al.} \cite{wangconsolidating}, Mishra {\it et al.} \cite{mishra2011theory} or Beloglazov {\it et al.} \cite{beloglazov2012energy}, who assumed models where consumption depended mainly on CPU and linearly on its utilization, proposing bin-packing-like algorithms to reduce power consumption.
Other works like the ones from Andrews {\it et al.} \cite{andrewsDiseconomies} or Irani {\it et al.} \cite{irani2007algorithms} proposed non-linear models, claiming that energy could 
be saved by running processes at the lowest possible speed.

Moving to the empirical field, we first classify works in two different groups, those who consider the effect of  frequency on their analysis and those who do not consider it. 
We start with those not considering frequency. In this category we find articles proposing models where server components follow a linear behavior 
like~\cite{krishnan2011vm, liu2011evaluating, vasan2010worth} or more complex ones, like in ~\cite{basmadjian2011methodology, economou2006full, lewis2008run}. In ~\cite{liu2011evaluating}, Liu {\it et al.} propose a simple linear model and evaluate different hardware configurations and types of workloads by varying the number of available cores, the available memory, and considering also the contribution of other components such as disks. 
%They depending linearly on the utilization of its components although did not give much detail about it as it was not the main focus of its work.  
Vasan {\it et al.}~\cite{vasan2010worth} monitored multiple servers on a datacenter as well as the power consumption of several of the internal elements of a server. However, they considered that the behavior of this server could be approximated by a model based only on  CPU utilization. Similarly, Krishnan {\it et al.}~\cite{krishnan2011vm} explored the 
feasibility of lightweight virtual machine power metering methods and examined the contribution of some of the elements that consume power in a server like CPU, memory and disks. 
Their model depends linearly on each of these components. In \cite{economou2006full}, Economou {\it et al.} proposed a non-intrusive method for modeling full-system 
power consumption by stressing its components with different workloads. Their resulting model is also linear on the utilization of its components. Finally, Lewis {\it et al.}~\cite{lewis2008run} and Basmasjian {\it et al.}~\cite{basmadjian2011methodology} presented much more complex models which, apart from the contribution of different components of the server, considered extra parameters like temperature and cache misses as well as multiple cores. In particular, Lewis {\it et al.}~\cite{lewis2008run} reported also an extensive study on the behavior of reading and writing operations in hard disk and solid state drives. 
In contrast, we show that linear models are not accurate and we complement the existing studies by showing the effect of different block sizes and frequencies, e.g., on network and individual read or write operations.

Now we move to the works which also considered frequency in their analysis. Miyoshi {\it et al.}~\cite{miyoshi2002critical} analyzed the runtime effects of frequency scaling on 
power and energy. 
% a part from introducing the \emph{critical power slope}\footnote{It tells us whether or not it is energy efficient to run at a higher performance state to complete a work and go into idle state} concept. 
Brihi {\it et al.}~\cite{brihi2013dynamic} presented an exhaustive study of DVFS using a \texttt{cpufrequtils} as we do. Main differences with our work were that they studied four different power management policies under DVFS and centered their study on the relationship between CPU utilization and power consumption. 
However, they also present interesting results about disk consumption that match partially our results, showing a flat consumption in reading operations 
and variations in the writing ones that they attribute to the size of the files being written. Although it was not the main objective of their work, Raghavendra {\it et al.}~\cite{raghavendra2008no} performed a per-frequency and core CPU power characterization of two different blade servers. However, they claimed that CPU power 
depends linearly on its utilization. The main difference with our analysis is that we consider that the load supported by a server increases with the number of active cores 
and, hence, this load should not be represented in percentage. Gandhi {\it et al.}~\cite{gandhi2009optimal} published a preliminary analysis of power consumption 
versus frequency, based on DVFS and DFS and gave some intuition about the non-linearity of this relation. However, our analysis is more complete as we present
a per-component analysis as well as enter into deeper details on the power versus frequency analysis.

%We conclude with some works that also consider frequency but do not model the power consumption of a server. First of them, the work from Le Sueur {\it et al.}~\cite{le2010dynamic} presents an analysis of the evolution of the effectiveness of DVFS and how it is reduced in the newest and most optimized servers. They show that DVSF might be soon obsoleted by the adoption of ultra low power sleep modes. Ge {\it et al.} proposed PowerPack \cite{ge2010powerpack}, a framework that includes a set of toolkits to perform an exhaustive profile of the power consumption of servers and its components. Their analysis is centered in showing the contribution of multicore system to the efficiency of several applications and, hence, no power consumption characterization is presented. Finally, Basmadjian {\it et al.}~\cite{basmadjian2012evaluating} published an in deep analysis of the components of a processor and its contribution to the power consumption of the CPU, shedding some light on the behavior of multicore servers. Some of their conclusions are very relevant for this work, as they show, for instance, that the power consumption of multiple cores performing parallel computations is not equal to the sum of the power of each of those active cores.

\section{Conclusions}
\label{sec:conclusions}

In this work we have reported our measurement-based characterization of energy and power consumption in a server. 
We have exhaustively measured the power consumed by CPU, disk, and NIC under different configurations, identifying the optimal operational levels, 
which usually do not correspond to the static system configurations commonly adopted. 
We found that, besides the {\it baseline component}, which does not changes significantly with the operational parameters, the CPU has the largest impact on energy consumption
among all the three components. We observe that CPU consumption is neither linear nor concave with the load. Disk I/O is the second larger contributor to power consumption, although performance changes sensibly with the I/O block size used by the applications. Finally, the NIC activity is responsible for a small but not negligible fraction of power consumption, which scales almost linearly with the network transmission rate.  
In general, most of the energy/power performance figures do not scale linearly with the utilization, in contrast to what is commonly assumed in the literature. 
We have then shown how to predict and optimize the energy consumed by an application via a concrete example using network 
activity plus pagerank computation in Hadoop. Our model achieves very accurate energy estimates, within $7\%$ or less from  the measured total power consumption.

%\input{appendix}

%COMMENTED BY ANGELOS ON 27/12/2013 
%\section{Acknowledgments}
%This research was supported in
%part by the Spanish MICINN grant TEC2011-29688-C02-01.
%COMMENTED BY ANGELOS ON 27/12/2013 

%\vspace{-1mm}
{\small
\bibliographystyle{acm}
\bibliography{biblio}

\begin{thebibliography}{10}

\bibitem{andrewsDiseconomies}
{\sc Andrews, M., Antonakopoulos, S., and Zhang, L.}
\newblock Minimum-cost network design with (dis)economies of scale.
\newblock In {\em IEEE FOCS\/} (2010), pp.~585--592.

\bibitem{basmadjian2011methodology}
{\sc Basmadjian, R., Ali, N., Niedermeier, F., de~Meer, H., and Giuliani, G.}
\newblock A methodology to predict the power consumption of servers in data
  centres.
\newblock In {\em ACM e-Energy\/} (2011), pp.~1--10.

\bibitem{beloglazov2012energy}
{\sc Beloglazov, A., Abawajy, J., and Buyya, R.}
\newblock Energy-aware resource allocation heuristics for efficient management
  of data centers for cloud computing.
\newblock {\em Future Generation Computer Systems 28}, 5 (2012), 755--768.

\bibitem{brihi2013dynamic}
{\sc Brihi, A., and Dargie, W.}
\newblock Dynamic voltage and frequency scaling in multimedia servers.
\newblock In {\em IEEE AINA\/} (2013).

\bibitem{economou2006full}
{\sc Economou, D., Rivoire, S., Kozyrakis, C., and Ranganathan, P.}
\newblock Full-system power analysis and modeling for server environments.
\newblock In {\em Proceedings of Workshop on Modeling, Benchmarking, and
  Simulation\/} (2006), pp.~70--77.

\bibitem{gandhi2009optimal}
{\sc Gandhi, A., Harchol-Balter, M., Das, R., and Lefurgy, C.}
\newblock Optimal power allocation in server farms.
\newblock In {\em ACM SIGMETRICS\/} (2009), pp.~157--168.

\bibitem{garcia2012energy}
{\sc Garcia-Saavedra, A., Serrano, P., Banchs, A., and Bianchi, G.}
\newblock Energy consumption anatomy of 802.11 devices and its implication on
  modeling and design.
\newblock In {\em ACM CoNEXT\/} (2012), pp.~169--180.

\bibitem{lambert}
{\sc Heddeghem, W.~V., Lambert, S., Lannoo, B., Colle, D., Pickavet, M., and
  Demeester, P.}
\newblock Trends in worldwide {ICT} electricity consumption from 2007 to 2012.
\newblock {\em Computer Communications\/} (Submitted).

\bibitem{802.3az}
{\sc {IEEE Std. 802.3az}}.
\newblock {Energy Efficient Ethernet}, 2010.

\bibitem{irani2007algorithms}
{\sc Irani, S., Shukla, S., and Gupta, R.}
\newblock Algorithms for power savings.
\newblock {\em ACM TALG 3}, 4 (2007), 41.

\bibitem{krishnan2011vm}
{\sc Krishnan, B., Amur, H., Gavrilovska, A., and Schwan, K.}
\newblock {VM} power metering: feasibility and challenges.
\newblock {\em ACM SIGMETRICS Performance Evaluation Review 38}, 3 (2011),
  56--60.

\bibitem{kusic2009power}
{\sc Kusic, D., Kephart, J.~O., Hanson, J.~E., Kandasamy, N., and Jiang, G.}
\newblock Power and performance management of virtualized computing
  environments via lookahead control.
\newblock {\em Cluster computing 12}, 1 (2009), 1--15.

\bibitem{lewis2008run}
{\sc Lewis, A.~W., Ghosh, S., and Tzeng, N.-F.}
\newblock Run-time energy consumption estimation based on workload in server
  systems.
\newblock {\em HotPower'08\/} (2008), 17--21.

\bibitem{liu2011evaluating}
{\sc Liu, C., Huang, J., Cao, Q., Wan, S., and Xie, C.}
\newblock Evaluating energy and performance for server-class hardware
  configurations.
\newblock In {\em IEEE NAS\/} (2011), pp.~339--347.

\bibitem{mishra2011theory}
{\sc Mishra, M., and Sahoo, A.}
\newblock On theory of vm placement: Anomalies in existing methodologies and
  their mitigation using a novel vector based approach.
\newblock In {\em IEEE CLOUD\/} (2011), pp.~275--282.

\bibitem{miyoshi2002critical}
{\sc Miyoshi, A., Lefurgy, C., Van~Hensbergen, E., Rajamony, R., and Rajkumar,
  R.}
\newblock Critical power slope: understanding the runtime effects of frequency
  scaling.
\newblock In {\em ACM ICS'02\/} (2002), pp.~35--44.

\bibitem{moore2005making}
{\sc Moore, J.~D., Chase, J.~S., Ranganathan, P., and Sharma, R.~K.}
\newblock Making scheduling ``cool'': Temperature-aware workload placement in
  data centers.
\newblock In {\em USENIX annual technical conference, General Track\/} (2005),
  pp.~61--75.

\bibitem{raghavendra2008no}
{\sc Raghavendra, R., Ranganathan, P., Talwar, V., Wang, Z., and Zhu, X.}
\newblock No power struggles: Coordinated multi-level power management for the
  data center.
\newblock In {\em ACM SIGARCH Computer Architecture News\/} (2008), vol.~36,
  ACM, pp.~48--59.

\bibitem{vasan2010worth}
{\sc Vasan, A., Sivasubramaniam, A., Shimpi, V., Sivabalan, T., and Subbiah,
  R.}
\newblock {Worth their Watts? - An empirical study of datacenter servers}.
\newblock In {\em IEEE HPCA\/} (2010), pp.~1--10.

\bibitem{wangconsolidating}
{\sc Wang, M., Meng, X., and Zhang, L.}
\newblock Consolidating virtual machines with dynamic bandwidth demand in data
  centers.
\newblock In {\em IEEE INFOCOM\/} (2011), pp.~71--75.

\bibitem{weiser1996scheduling}
{\sc Weiser, M., Welch, B., Demers, A., and Shenker, S.}
\newblock Scheduling for reduced {CPU} energy.
\newblock In {\em Mobile Computing}. Springer, 1996, pp.~449--471.

\end{thebibliography}
}

%\newpage
%\input{appendix}
\vfill
\end{document}